\documentclass[aps,showpacs,pra,superscriptaddress,]{revtex4}
\usepackage{amsmath}
\usepackage{amsfonts}
\usepackage{amssymb}
\usepackage{bm}
\usepackage{graphicx}

\begin{document}

\title{Chirped periodic and solitary waves in nonlinear negative index
materials}
\author{Houria Triki}
\affiliation{Radiation Physics Laboratory, Department of Physics, Faculty of Sciences,
Badji Mokhtar University, P. O. Box 12, 23000 Annaba, Algeria }
\author{ Vladimir I. Kruglov}
\affiliation{Centre for Engineering Quantum Systems, School of Mathematics and Physics,
The University of Queensland, Brisbane, Queensland 4072, Australia}

\begin{abstract}
Propagation of ultrashort pulses at least a few tens of optical cycles in
duration through a negative index material is investigated theoretically
based on the generalized nonlinear Schr\"{o}dinger equation with
pseudo-quintic nonlinearity and self-steepening effect. Novel periodic waves
of different forms are shown to exist in the system in the presence of
higher-order effects. It is found that such periodic structures exhibit an
interesting chirping property which depends on the light field intensity.
The nonlinearity in pulse chirp is found to be caused by the presence of
self-steepening effect in the negative index medium. The solutions also comprise dark, bright, and kink solitary waves. Stability of the solitary wave solutions are proved analytically using the theory of nonlinear dispersive waves. The stability of the solutions is numerically studied under finite initial perturbations.
\end{abstract}

\maketitle
\affiliation{$^{1}${\small Radiation Physics Laboratory, Department of Physics, Faculty
of Sciences, Badji Mokhtar University, P. O. Box 12, 23000 Annaba, Algeria}\\
$^{2}${\small Centre for Engineering Quantum Systems, School of Mathematics
and Physics, The University of Queensland, Brisbane, Queensland 4072,
Australia}}

\section{Introduction}

Propagation of ultrashort light pulses in negative index materials (NIMs)
has attracted considerable attention recently \cite{PLi}-\cite{Shen}. Such
new materials are composed of a regular array of unit cells whose size is
usually much smaller than the wavelengths of propagating electromagnetic
waves \cite{PLi}. Their remarkable property of having an antiparallel phase
velocity and Poynting vector \cite{Pendry}-\cite{Parazzoli}, makes them the
most potential candidates for stable soliton and other nonlinear phenomena 
\cite{Xiang1}. Compared with an ordinary material which is a medium
possessing a constant permeability and positive refraction, a NIM has a
dispersive permeability and a negative refraction. After the Veselago's
prediction \cite{Veselago} and the experimental demonstration of NIMs \cite%
{Pendry1,Smith1,Smith2}, interest has been aroused in the study of the
potential applications of such materials, which display properties beyond
those available in naturally occurring materials. Further, experimental
results have been recently obtained with studying nonlinear phenomena in
NIMs, including parametric amplification \cite{Aguanno}, modulation
instability \cite{Popov}, second-harmonic generation \cite{Wen}, and soliton
propagation \cite{Ban,Boardman}.

Theoretically, Scalora et al. \cite{Scalora} have introduced a generalized
nonlinear Schr\"{o}dinger equation (NLSE) for a dispersive dielectric
susceptibility and permeability appropriate for describing ultrashort pulse
propagation in NIMs. This new NLSE model includes the contribution of
higher-order effects such as pseudo-quintic nonlinearity and self-steepening
effect with parameters directly related to the magnetic properties of the
material. Development of such generalized NLSE was a great advance in the
study of ultrashort electromagnetic pulses through NIMs. Subsequently, Zhang
and Yi \cite{ZY} obtained the exact chirped soliton solutions of this newly
derived equation by employing the variable parametric method. Later,
Marklund et al. \cite{Marklund} also examined the modulational instability
and localization of an ultrashort electromagnetic pulse that is governed by
this generalized NLSE. Recently, Yang et al. \cite{Yang} have discussed the
existence of quasi-solitons within the framework of this generalized NLSE
under the very specific condition of the vanishing self-steepening
nonlinearity. As a particular result, they have obtained three types of
exact bright, dark, bright-grey quasi-soliton solutions by employing the
ansatz method. But with the increasing intensity of the optical field and
further shortening of the pulses up to the femtosecond duration ($<100~%
\mathrm{fs}$), the self-steepening process becomes increasingly more
important in optical systems \cite{Han}. Such higher-order nonlinear effect
is due to the intensity dependence of group velocity, which gives the
optical pulse a very narrow width during the propagation \cite{Porsezian1}.
It is worth noting that the presence of self-steepening process strongly
affects the propagation of ultrashort pulses in nonlinear media. Thus,
taking into account the contribution of this higher-order effect is relevant
to understand experimental results and complement the previous studies. It
should be mentioned that the addition of a higher-order nonlinear term to a
given nonlinear evolution equation changes drastically it's integrability
properties.

Regarding the application of solitons, how to control the balance
between nonlinear and dispersion effects becomes a significant subject in
ultrafast optics \cite{Liu1}. In this context, new materials that can be
successfully used for the generation of ultrashort pulses have been recently
demonstrated \cite{Liu2,Liu3}. In addition, heterostructure materials having
excellent optical properties in photonic device applications have been
recently fabricated by employing the magnetron sputtering technique \cite%
{Liu4}. Theoretically, nonlinear control based on different dispersion and
nonlinearity for third-order NLSE has been recently studied \cite{Liu5}. It
is of interest to note that the investigation of soliton dynamics in the
presence of higher-order effects has been also expanded into optical systems
which are modeled by the complex cubic-quintic Ginzburg-Landau equation \cite{Liu6,Liu7}.

Interest in nonlinear periodic waves has grown rapidly in recent years due
to their potential applications in many fields of physics \cite{Shamy}-\cite%
{Lai}. The study of such structures has made a remarkable progress because
of their use in the nonlinear transport phenomenon \cite{Shamy}. Note that
cnoidal waves based on elliptic functions like $\mathrm{cn,}$ $\mathrm{sn}$
as well as $\mathrm{dn}$ are exact solutions in the form of periodic arrays
of pulses. In optics, cnoidal (periodic) waves play a significant role in
the analysis of the data transmission in fiber-optic telecommunications
links \cite{D}. Due to their structural stability with respect to the small
input profile perturbations and collisions \cite{Petnikova}, these
propagating waves serve as a model of pulse train propagation in optics
fibers\ \cite{D}. In this context, we recently demonstrated the formation of
periodic waves and localized pulses in an optical fiber medium under the
influence of third-order dispersion and self-steepening effect, which can
stably propagate through the system with a Kerr nonlinear response \cite%
{Vladimir}. Although periodic waves have been widely studied in physical
systems such as nonlinear optical fibers \cite{Vladimir,Chow,Chow1} and
Bose-Einstein condensates \cite{FKh,Chow2}, their investigation in NIMs has
not been widespread. In addition, most previous studies on propagating
periodic waves have been limited to the linearly chirped and chirp-free
periodic structures \cite{Vladimir}-\cite{Kruglov}. An issue of great
current interest is the search for periodic waves that are characterized by
an enhanced nonlinearity in pulse chirp. From a practical stand point,
chirped pulses are particularly useful in the design of optical devices such
as fiber-optic amplifiers, optical pulse compressors and solitary wave based
communication links \cite{Kruglov,Desaix}. In this paper, we study the
formation of periodic waves in a NIM exhibiting pseudo-quintic nonlinearity
and self-steepening effects. We find the periodic waves with these
higher-order effects have a wide variety of functional forms with properties
different from the ones existing in optical Kerr media. Importantly, the
structures we obtained exhibit explicitly a nonlinear chirp that arises from
the self-steepening process, and thus describe physically important
applications as the compression and amplification of light pulses in
negative index materials. Conditions for their existence as well as
stability aspects of such privileged exact solutions are also investigated.

The paper is arranged as follows. In Sec. II, we briefly present the
generalized NLSE describing ultrashort optical pulse propagation in a NIM
and derive the nonlinear equation that governs the evolution of the light
field intensity in the system. In Sec. III, we present results of novel
chirped periodic solutions of the model and the nonlinear chirp associated
with each of these structures. The solitary-wave (long-wave) limit of the
periodic solutions is also considered here, which has given rise to
localized pulses of bright, dark, gray, kink, and anti-kink type. In Sec.
IV, the analytical stability analysis of chirped periodic and solitary wave
solutions based on the theory of optical nonlinear dispersive waves is
presented. In Sec. V, the the stability of the solutions is discussed
numerically. The paper is concluded by Sec. VI (where, in particular, we
present applications of the obtained periodic solutions in other systems
such as nonlinear optical fibers).

\section{Traveling waves of generalized nonlinear Schr\"{o}dinger equation}

The governing equation for ultrashort pulse propagation in a negative index
material with pseudo-quintic nonlinearity and self-steepening effect is
given by the generalized NLSE \cite{Scalora}:%
\begin{equation}
i\frac{\partial \psi }{\partial z}+\frac{\sigma }{2}\frac{\partial ^{2}\psi 
}{\partial t^{2}}+\rho \left\vert \psi \right\vert ^{2}\psi -\epsilon
\left\vert \psi \right\vert ^{4}\psi -i\nu \frac{\partial }{\partial t}%
(\left\vert \psi \right\vert ^{2}\psi )=0,  \label{1}
\end{equation}%
where $\psi \left( z,t\right) $ is the complex envelope of the electric
field, with $z$ and $t$ the normalized propagation distance and time,
respectively. The coefficient $\sigma =\left( 1/\beta n\right) \left[
1/V_{g}^{2}-\alpha \gamma -\beta \left( \varepsilon \gamma ^{\prime }+\mu
\alpha ^{\prime }\right) /4\pi \right] $ denotes the group-velocity
dispersion (GVD) parameter, $\rho =\beta \mu \chi ^{(3)}/2n$ and $\epsilon
=\beta \mu ^{2}(\chi ^{(3)})^{2}/8n^{3}$ are the cubic and pseudo-quintic
nonlinearity coefficients, respectively, and $\nu =\chi ^{(3)}[\mu
/2V_{g}n^{2}-(\gamma +\mu )/2n]$ accounts for the pulse self-steepening
effect. Here $\beta =2\pi \widetilde{\omega }=2\pi \omega /\omega _{p}$
where $\omega _{p}$ is the electric plasma frequency, $\chi ^{(3)}$ is the
third order susceptibility of the medium, $\mu $ is the effective magnetic
permeability, $n$ is the refractive index, and $V_{g}=2n/\left( \varepsilon
\gamma +\mu \alpha \right) $ is the group velocity of the pulse. Also $%
\alpha =\partial \left[ \widetilde{\omega }\varepsilon (\widetilde{\omega })%
\right] /\partial \widetilde{\omega }$, $\alpha ^{\prime }=\partial ^{2}%
\left[ \widetilde{\omega }\varepsilon (\widetilde{\omega })\right] /\partial 
\widetilde{\omega }^{2},$ $\gamma =\partial \left[ \widetilde{\omega }\mu (%
\widetilde{\omega })\right] /\partial \widetilde{\omega }$, $\gamma ^{\prime
}=\partial ^{2}\left[ \widetilde{\omega }\mu (\widetilde{\omega })\right]
/\partial \widetilde{\omega }^{2},$ where $\varepsilon $ and $\widetilde{%
\omega }$ denote dielectric susceptibility and normalized frequency,
respectively. In what follows, we investigate the formation
conditions and propagation properties of periodic and solitary waves and
take the same values of the model parameters used in Ref. \cite{Yang}. Such
realistic parameter values are chosen depending on which ratio of $\omega
_{m}/\omega _{p}$ is chosen to determine the engineered size of the
constituents of NIM structures, the split-ring resonators, and so on (with $\omega _{m}$ is the magnetic plasma frequency). One should note
that for categorizing the generalized NLSE to describe the behavior of
negative refraction, it is further correlated with the lossless Drude model
delineating the frequency dispersion with the permittivity and permeability
of the form \cite{Scalora}: $\varepsilon (\widetilde{\omega })=1-(1/(%
\widetilde{\omega }^{2}+i\widetilde{\omega }\gamma ))$ and $\mu (%
\widetilde{\omega })=1-((\omega _{m}^{2}/\omega _{p}^{2})/(\widetilde{\omega 
}^{2}+i\widetilde{\omega }\gamma ))$. Therefore, the
refractive index of the medium defined by $n(\widetilde{\omega })=\pm \sqrt{%
\varepsilon (\widetilde{\omega })\mu (\widetilde{\omega })}$ is
negative for $\widetilde{\omega }<0.8$ and positive for $%
\widetilde{\omega }>1.0$ if considering the typical value $\omega
_{m}/\omega _{p}=0.8$ \cite{Yang,Min}. Following the
expressions of the preceding parameters, for self-focusing nonlinearity, $%
\rho $ $>0$ and $\epsilon <0$ in the
negative-index region $\widetilde{\omega }<0.8$, $\rho $
$>0$ and $\epsilon >0$ in the positive-index region $%
\widetilde{\omega }>1.0$; while for self-defocusing nonlinearity, $%
\rho $ $<0$ and $\epsilon <0$ in the negative
index region and $\rho $ $<0$ and $\epsilon >0$ in the positive index region \cite{Yang}. In addition, the sign of the GVD
coefficient $\sigma $ can be positive or negative depending on the
particular choice of parameters, and self-steepening characterizes the front
of the pulse, unlike in the case of ordinary materials \cite{Scalora}. 
One should also note that all the model parameters in the
generalized NLSE (\ref{1}) are function of the normalized frequency $%
\widetilde{\omega }$, whose specific dependence curves is depicted
in figure 1 of Ref. \cite{Tsitsas}. As stated in Refs. \cite%
{PLi,Scalora,Wen1}, the model parameters directly expressed by material
permittivity and permeability $\varepsilon (\widetilde{\omega })$
and $\mu (\widetilde{\omega })$ can be tailored by engineering the
unit-cell structure of \ negative index materials, which implies more
possibilities for the existence of wider classes of solitary waves \cite%
{Scalora}. 

Equation (\ref{1}) for $\epsilon =0$ reduces to the derivative NLSE that
describes the optical soliton propagation in the presence of Kerr dispersion 
\cite{Tzoar,Anderson}. As mentioned previously, exact bright, dark,
bright-grey quasi-soliton solutions of Eq. (\ref{1}) have been obtained
under the special parametric condition $\nu =0$ in \cite{Yang}. Here instead
we shall analyze the existence of nonlinearly chirped periodic and solitary
waves taking into account the influence of all effects on short-pulse
propagation in the negative refractive index material.

In order to\ obtain the exact traveling-wave solutions of generalized NLSE (%
\ref{1}), we consider a solution of the form, 
\begin{equation}
\psi (z,t)=u(x)\exp [i\left( \kappa z-\delta t\right) +i\phi (x)],  \label{2}
\end{equation}%
where $u(x)$ and $\phi (x)$ are real functions depending on the variable $%
x=t-qz$, and $q=v^{-1}$ is the inverse velocity. Here $\kappa $ and $\delta $
are the corresponding real parameters describing the wave number and
frequency shift. Equations (\ref{1}) and (\ref{2}) lead to the following
system of the ordinary differential equations, 
\begin{equation}
\frac{\sigma }{2}\left( u\frac{d^{2}\phi }{dx^{2}}+2\frac{d\phi }{dx}\frac{du%
}{dx}\right) -\left( q+\sigma \delta \right) \frac{du}{dx}-3\nu u^{2}\frac{du%
}{dx}=0,  \label{3}
\end{equation}%
\begin{eqnarray}
\frac{\sigma }{2}\frac{d^{2}u}{dx^{2}}-\left( \kappa -q\frac{d\phi }{dx}%
\right) u-\frac{\sigma }{2}\left( \frac{d\phi }{dx}-\delta \right) ^{2}u 
\notag \\
\noalign{\vskip3pt}+\nu\frac{d\phi }{dx}u^{3}+\left( \rho -\nu \delta
\right) u^{3}-\epsilon u^{5}=0.  \label{4}
\end{eqnarray}%
Multiplying Eq. (\ref{3}) by the function $u(x)$ and integrating leads to
the following equation, 
\begin{equation}
\sigma u^{2}\frac{d\phi }{dx}-(q+\sigma \delta )u^{2}-\frac{3}{2}\nu
u^{4}=K_{0},  \label{5}
\end{equation}%
where $K_{0}$ is the integration constant. We choose below $K_{0}=0$ then
Eq. (\ref{5}) yields, 
\begin{equation}
\frac{d\phi }{dx}=\frac{q+\sigma \delta }{\sigma }+\frac{3\nu }{2\sigma }%
u^{2}.  \label{6}
\end{equation}%
The corresponding chirp $\Delta \omega $ defined as $\Delta \omega
=-\partial \left[ \kappa z-\delta t+\phi (x)\right] /\partial t$ is given by%
\begin{equation}
\Delta \omega =-\frac{q}{\sigma }-\frac{3\nu }{2\sigma }u^{2}.  \label{7}
\end{equation}%
This result shows that the chirp is strongly dependent on the
self-steepening parameter $\nu $ and varies with the wave intensity $%
\left\vert \psi \right\vert ^{2}=u^{2}$. Accordingly, a nontrivial nature of
the phase will take place, thus making the envelope pulses nonlinearly
chirped. It is clear that this chirping property disappears in the absence
of self-steepening effect (i.e., $\nu =0$), where only the unchirped pulses
are allowed to exist in the negative index medium in the presence of the
pseudo-quintic nonlinearity.

Further insertion of (\ref{6}) into (\ref{4}) gives to the following
nonlinear ordinary differential equation, 
\begin{equation}
\frac{d^{2}u}{dx^{2}}+au+bu^{3}+cu^{5}=0,  \label{8}
\end{equation}%
with the parameters $a,$ $b$ and $c$ given by%
\begin{equation}
a=\frac{q^{2}+2\sigma \left( \delta q-\kappa \right) }{\sigma ^{2}},~~~~b=%
\frac{2(\rho \sigma +\nu q)}{\sigma ^{2}},~~~~c=\frac{3\nu ^{2}-8\sigma
\epsilon }{4\sigma ^{2}}.  \label{9}
\end{equation}

\noindent Integration of Eq. (\ref{8}) yields the first order nonlinear
differential equation, 
\begin{equation}
\left( \frac{du}{dx}\right) ^{2}+au^{2}+\frac{1}{2}bu^{4}+\frac{1}{3}%
cu^{6}=h_{0},  \label{10}
\end{equation}%
where $h_{0}$ is an integration constant. We define new function $%
Y(x)=u^{2}(x)$ which transforms Eq. (\ref{10}) to the form, 
\begin{equation}
\frac{1}{4\alpha _{0}}\left( \frac{dY}{dx}\right) ^{2}=Y^{4}+\alpha
_{3}Y^{3}+\alpha _{2}Y^{2}+\alpha _{1}Y,  \label{11}
\end{equation}%
with $\alpha _{0}=-c/3$, $\alpha _{3}=3b/2c$, $\alpha _{2}=3a/c$ and $\alpha
_{1}=-3h_{0}/c$. We also define the polynomial $f(Y)$: 
\begin{equation}
f(Y)=Y^{4}+\alpha _{3}Y^{3}+\alpha _{2}Y^{2}+\alpha
_{1}Y=\prod_{i=1}^{4}(Y-Y_{i}),  \label{12}
\end{equation}%
where $Y_{i}$ are the roots of this polynomial. The previous equation (\ref%
{11}) is a nonlinear ordinary differential equation, which governs the
evolution of the wave intensity as it propagates through the negative
refractive index material. It gives a complete description of the envelope
dynamics in NIMs under the influence of higher-order effects such as
pseudoquintic nonlinearity and self-steepening effect. The general solution
of Eq. (\ref{11}) has the form, 
\begin{equation}
\pm \int \frac{dY}{\sqrt{\mathrm{sgn}(\alpha _{0})(Y^{4}+\alpha
_{3}Y^{3}+\alpha _{2}Y^{2}+\alpha _{1}Y)}}=2\sqrt{|\alpha _{0}|}(x-\eta ),
\label{13}
\end{equation}%
where $\mathrm{sgn}(\alpha _{0})=\alpha _{0}/|\alpha _{0}|$ for $\alpha
_{0}\neq 0$, and $\eta $ is an integration constant. We consider in the
following section different cases which yield the periodic bounded and
solitary wave solutions of the generalized NLSE (\ref{1}).

\section{ Chirped periodic and solitary wave solutions}

We consider here the particular case with real roots for equation $f(Y)=0$
and $\alpha _{0}=-c/3>0$. Let all roots are different then we can select
them as $Y_{1}>Y_{2}>Y_{3}>Y_{4}$. We consider the case when function $Y(x)$
belongs to the next interval $Y_{2}\geq Y\geq Y_{3}$. The integral in Eq. (%
\ref{13}) for real roots $Y_{i}$ with $Y_{1}>Y_{2}>Y_{3}>Y_{4}$ and $%
Y_{2}\geq Y\geq Y_{3}$ can be written in the next form, 
\begin{equation}
\int_{Y}^{Y_{2}}\frac{dY^{\prime }}{\sqrt{(Y^{\prime }-Y_{1})(Y^{\prime
}-Y_{2})(Y^{\prime }-Y_{3})(Y^{\prime }-Y_{4})}}=\frac{2F(s;k)}{\sqrt{%
(Y_{1}-Y_{3})(Y_{2}-Y_{4})}},  \label{14}
\end{equation}%
where the parameters $s$ and $k$ are given by 
\begin{equation}
s=\left( \frac{(Y_{1}-Y_{3})(Y_{2}-Y)}{(Y_{2}-Y_{3})(Y_{1}-Y)}\right)
^{1/2},~~~~k=\left( \frac{(Y_{2}-Y_{3})(Y_{1}-Y_{4})}{%
(Y_{1}-Y_{3})(Y_{2}-Y_{4})}\right) ^{1/2}.  \label{15}
\end{equation}%
The function $F(s;k)$ is Jacoby form of the elliptic integral of the first
kind, 
\begin{equation}
F(s;k)=\int_{0}^{s}\frac{dx}{\sqrt{(1-x^{2})(1-k^{2}x^{2})}}.  \label{16}
\end{equation}%
Equations (\ref{13}) and (\ref{14}) lead the following relation, 
\begin{equation}
\frac{(Y_{1}-Y_{3})(Y_{2}-Y(x))}{(Y_{2}-Y_{3})(Y_{1}-Y(x))}=\mathrm{sn}%
^{2}(w(x-\eta ),k),  \label{17}
\end{equation}%
where $\mathrm{sn}(z,k)$ is the Jacoby elliptic function ($0\leq k\leq 1$)
and the parameter $w$ is given by 
\begin{equation}
w=\sqrt{\alpha _{0}(Y_{1}-Y_{3})(Y_{2}-Y_{4})},  \label{18}
\end{equation}%
with $\alpha _{0}>0$. Equation (\ref{17}) yields the periodic bounded
solution for the function $u(x)$ when the parameter $k$ in Eq. (\ref{15})
belongs to the interval $0<k<1$. This periodic bounded solutions are given
by 
\begin{equation}
u(x)=\pm \sqrt{\frac{Y_{2}-Y_{1}\Lambda \mathrm{sn}^{2}(w(x-\eta ),k)}{%
1-\Lambda \mathrm{sn}^{2}(w(x-\eta ),k)}},~~~~\Lambda =\frac{Y_{2}-Y_{3}}{%
Y_{1}-Y_{3}}.  \label{19}
\end{equation}%
We show below that this periodic solution reduces to soliton solution when
the parameter $k=1$. The period $T$ of this bounded periodic solution is $%
T=(2/w)K(k)$ where the function $K(k)$ is given by 
\begin{equation}
K(k)=\int_{0}^{1}\frac{dx}{\sqrt{(1-x^{2})(1-k^{2}x^{2})}}=F(1;k).
\label{20}
\end{equation}

The nonlinearity in pulse chirp is caused by the presence of self-steepening effect in the negative index medium. A rich variety of nonlinearly chirped solitary waves are obtained in the long-wave limit
of the derived periodic waves. The solutions contain dark, bright, and kink
solitary wave solutions, demonstrating the structural diversity of localized
waves in the material. We show below that the general solutions in Eqs. (\ref{13}) and (\ref{19}) lead to different periodic and solitary wave solutions of the generalized NLSE (\ref{1}). 

\begin{description}
\item[\textbf{1. Periodic bounded waves}] 
\end{description}

We have found a family of nonlinearly chirped periodic bounded solutions by Eqs. (\ref{2}) and (\ref{19}) as
\begin{equation}
\psi (z,t)=\pm \sqrt{\frac{Y_{2}-Y_{1}\Lambda \mathrm{sn}^{2}(w\xi ,k)}{%
1-\Lambda \mathrm{sn}^{2}(w\xi ,k)}}\exp [i\left( \kappa z-\delta t\right)
+i\phi (x)],  \label{21}
\end{equation}%
where $\xi =t-qz-\eta $, parameter $k$ ($0<k<1$) is given by Eq. (\ref{15}),
and parameter $w$ is defined by Eq. (\ref{18}) with $\alpha _{0}=-c/3>0$.
The phase $\phi (x)$ in this solution is following by Eq. (\ref{6}) as 
\begin{equation}
\phi (x)=\left( \frac{q}{\sigma }+\delta \right) (x-\eta )+\frac{3\nu }{%
2\sigma }\int_{\eta }^{x}u^{2}(x^{\prime })dx^{\prime }+\phi _{0},
\label{22}
\end{equation}

\noindent which shows that under the influence of self-steepening
nonlinearity, the propagating waveforms acquires an extra characteristic
phase $\phi (x)$ which depends on the squared amplitude of the complex
envelope and varies with the self-steepening parameter $\nu $. To be
specific, in the presence of self-steepening effect, the phase (\ref{22})
takes a nontrivial structure leading to chirped pulses. Obviously, in the
limit of $\nu =0$ (that is, at a vanishing self-steepening effect), the
phase distribution becomes a linear function of $\xi $.

Equation (\ref{15}) yields $k=1$ in the limiting case $Y_{3}=Y_{4}$. Note
that function $f(Y)$ has at least one root equal to zero. Thus in the
limiting case with $k=1$ we have $Y_{3}=Y_{4}=0$ and the parameter $\alpha
_{1}=0$. Hence the case with $k=1$ occurs when the integration constant $%
h_{0}=0$. In this limiting case, the periodical solution in Eq. (\ref{19})
reduces to soliton solution as 
\begin{equation}
u(x)=\pm\sqrt{\frac{Y_{2}[1-\mathrm{tanh}^{2}(w\xi)]}{1-(Y_{2}/Y_{1})\mathrm{%
tanh}^{2}(w\xi)}},  \label{23}
\end{equation}%
where $w=\sqrt{-(c/3)Y_{1}Y_{2}}$ and $Y_{1}$ and $Y_{2}$ are the roots of
equation $Y^{2}+\alpha _{3}Y+\alpha _{2}=0$. One can also write this
solution in the next equivalent form, 
\begin{equation}
u(x)=\frac{\pm\sqrt{Y_{1}}}{\sqrt{1+[(Y_{1}/Y_{2})-1]\mathrm{cosh}^{2}(w\xi)}%
},  \label{24}
\end{equation}%
where $Y_{1}>Y_{2}$. The roots $Y_{1}$ and $Y_{2}$ are given by 
\begin{equation}
Y_{1}=\frac{3b+\sqrt{3(3b^{2}-16ac)}}{4|c|},~~~~Y_{2}=\frac{3b-\sqrt{
3(3b^{2}-16ac)}}{4|c|}.  \label{25}
\end{equation}%
One can also write the soliton solution given in Eq. (\ref{24}) as 
\begin{equation}
u(x)=\frac{A}{\sqrt{1+B\mathrm{cosh}(w_{0}\xi)}},  \label{26}
\end{equation}%
where $\xi=t-qz-\eta$, and the parameters $A$, $B$, and $w_{0}$ are 
\begin{equation}
A=\pm\sqrt{\frac{2Y_{1}Y_{2}}{Y_{1}+Y_{2}}}=\pm2\sqrt{-\frac{a}{b}},
\label{27}
\end{equation}%
\begin{equation}
B=\frac{Y_{1}-Y_{2}}{Y_{1}+Y_{2}}=\sqrt{1-\frac{16ac}{3b^{2}}},  \label{28}
\end{equation}%
\begin{equation}
w_{0}=\sqrt{-(c/3)Y_{1}Y_{2}} =2\sqrt{-a},  \label{29}
\end{equation}%
with $a<0$, $b>0$, and $3b^{2}>16ac$.

\begin{description}
\item[\textbf{2. Bright solitary waves}] 
\end{description}

Substitution of (\ref{26}) into (\ref{2}) yields to the following exact
chirped bright solitary wave solution of Eq. (\ref{1}) as 
\begin{equation}
\psi (z,t)=\frac{A}{\sqrt{1+B\cosh (w_{0}\xi )}}\exp [i\left( \kappa
z-\delta t\right) +i\phi (x)],  \label{30}
\end{equation}%
with $a<0$, $b>0$, and $3b^{2}>16ac$. Here $\xi =t-qz-\eta $, and the
parameters $A$, $B$, and $w_{0}$ are given in Eqs. (\ref{27}), (\ref{28})
and (\ref{29}) respectively. The phase $\phi (x)$ is evaluated by Eqs. (\ref%
{22}) and (\ref{26}) as 
\begin{eqnarray}
\phi (x) = \frac{3\nu A^{2}}{\sigma w_{0}D}\mathrm{arctanh}\left( \frac{D}{%
1+B}~\mathrm{tanh}\left( \frac{1}{2}w_{0}(x-\eta )\right) \right)  \notag \\
\noalign{\vskip3pt}+\left( \frac{q}{\sigma }+\delta \right) (x-\eta
)+\phi_{0},~~~~~~~~~~~~~~~~~~  \label{31}
\end{eqnarray}%
where $\phi _{0}$ is integration constant and the parameter $D$ is 
\begin{equation}
D=\sqrt{1-B^{2}}=\sqrt{\frac{16ac}{3b^{2}}}.  \label{32}
\end{equation}%
Note that $D$ is real and $D\neq 0$ when $c<0$ because in this soliton
solution $a<0$.

\begin{figure}[tbp]
\includegraphics[width=\textwidth]{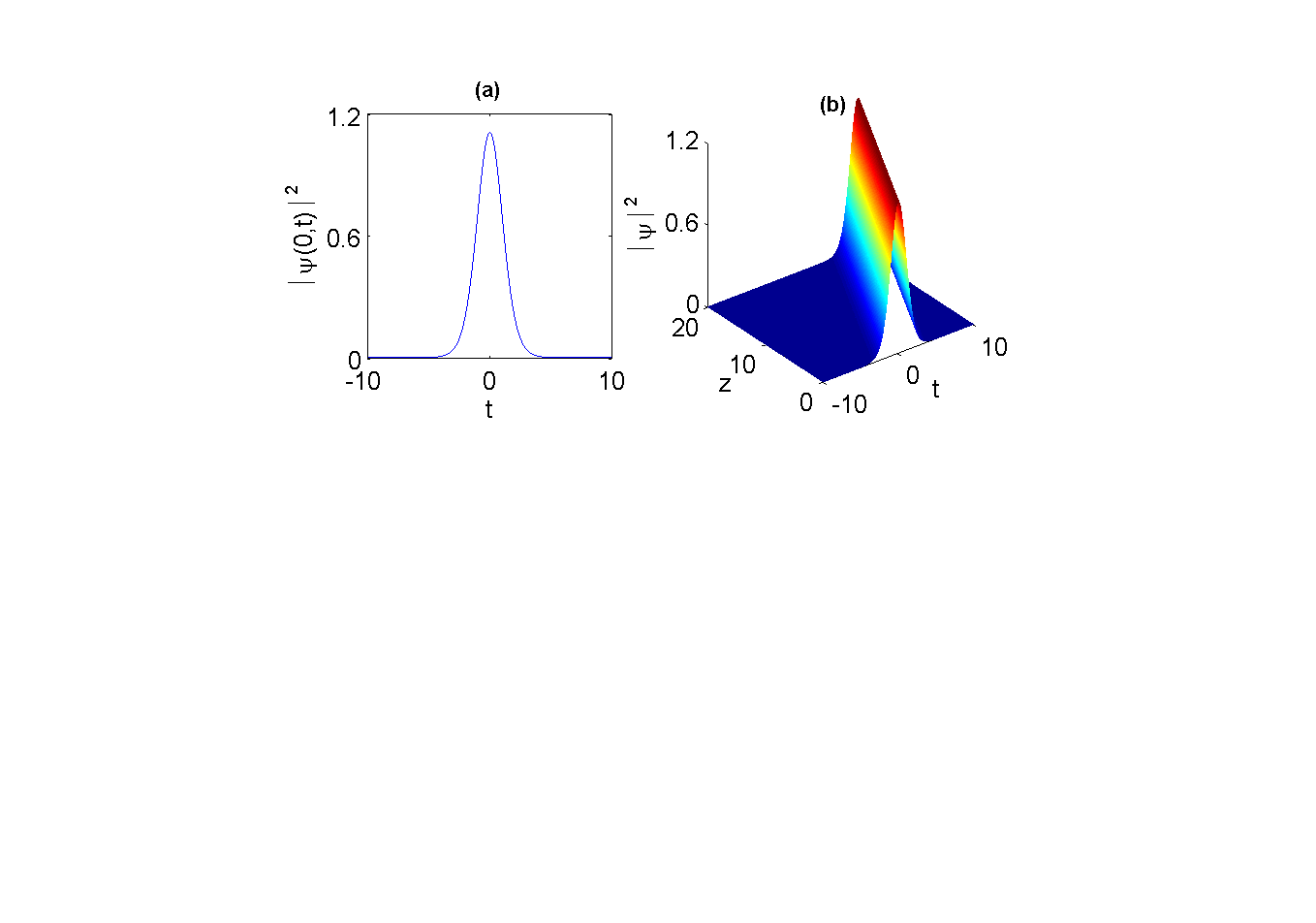} 
\caption{(a) Intensity profile of the chirped bright solitary wave $%
\left\vert \protect\psi (0,t)\right\vert ^{2}$ as a function of $t$ and its
(b) evolution as computed from Eq. (\ref{30}) for the values $\protect\sigma %
=-0.1990,$ $\protect\rho =-5.0265\times 10^{-10},$ $\protect\epsilon %
=3.3511\times 10^{-20},$ $\protect\nu =-6.2918\times 10^{-14},$ $q=0.5,$ $%
\protect\kappa =0.257,$ $\protect\eta =0,$ and $\protect\delta =1.82$. Here
the solitary wave intensity is normalized by $\left\vert \protect\psi %
(z,t)\right\vert ^{2}/A^{2}.$}
\label{FIG.1.}
\end{figure}

We have shown the solitary wave profile and its evolution in Figs. 1(a) and
1(b) respectively, for the parameter values \cite{Yang}: $\sigma =-0.1990,$ $%
\rho =-5.0265\times 10^{-10},$ $\epsilon =3.3511\times 10^{-20},$ and $\nu
=-6.2918\times 10^{-14}$. Other parameters are $q=0.5,$ $\kappa =0.257,$ $%
\eta =0,$ and $\delta =1.82$. It is clearly seen that this nonlinearly
chirped bright pulse propagate on a zero background in the negative index
medium in the presence of various physical effects. This demonstrated that
bright-type solitary waves that are characterized by a nonlinear chirp can
exist in the system and the formation of these structures is strongly
depends on all the material parameters.

\begin{description}
\item[\textbf{3. Periodic waves connected to bright solitary waves}] 
\end{description}

The soliton solution in Eq. (\ref{30}) is connected with the special
periodic solution when $a>0$, $b<0$, and $3b^{2}>16ac$. We have found such
an exact periodic wave solution of Eq. (\ref{8}) as $u(x)=A[1+B\cos
(w(x-\eta ))]^{-1/2}$. In this case, the chirped periodic solution of Eq. (%
\ref{1}) has the form, 
\begin{equation}
\psi (z,t)=\frac{A}{\sqrt{1+B\cos (w(x-\eta ))}}\exp [i\left( \kappa
z-\delta t\right) +i\phi (x)],  \label{33}
\end{equation}%
where $x=t-qz$, and the parameters $A$, $B$, and $w$ are given by 
\begin{equation}
A^{2}=-\frac{4a}{b},~~~~B=\pm \sqrt{1-\frac{16ac}{3b^{2}}},~~~~w=2\sqrt{a}.
\label{34}
\end{equation}%
Hence the periodic solution in Eq. (\ref{33}) is bounded when $a>0$, $b<0$, $%
3b^{2}>16ac$, and $|B|<1$. The phase $\phi (x)$ is evaluated by Eqs. (\ref%
{22}) and (\ref{33}) as 
\begin{eqnarray}
\phi (x) = \frac{3\nu A^{2}}{\sigma wD}\mathrm{arctan}\left( \frac{D}{1+B}~%
\mathrm{tan}\left( \frac{1}{2}w(x-\eta )\right) \right)  \notag \\
\noalign{\vskip3pt}+\left( \frac{q}{\sigma }+\delta \right) (x-\eta )+\phi
_{0},~~~~~~~~~~~~~~~~~~  \label{35}
\end{eqnarray}%
where $\phi _{0}$ is integration constant and $D=\sqrt{1-B^{2}}=\sqrt{%
16ac/3b^{2}}$. Note that $D$ is real and $D\neq 0$ when $c>0$ because in
this periodic solution $a>0$. 
\begin{figure}[tbp]
\includegraphics[width=\textwidth]{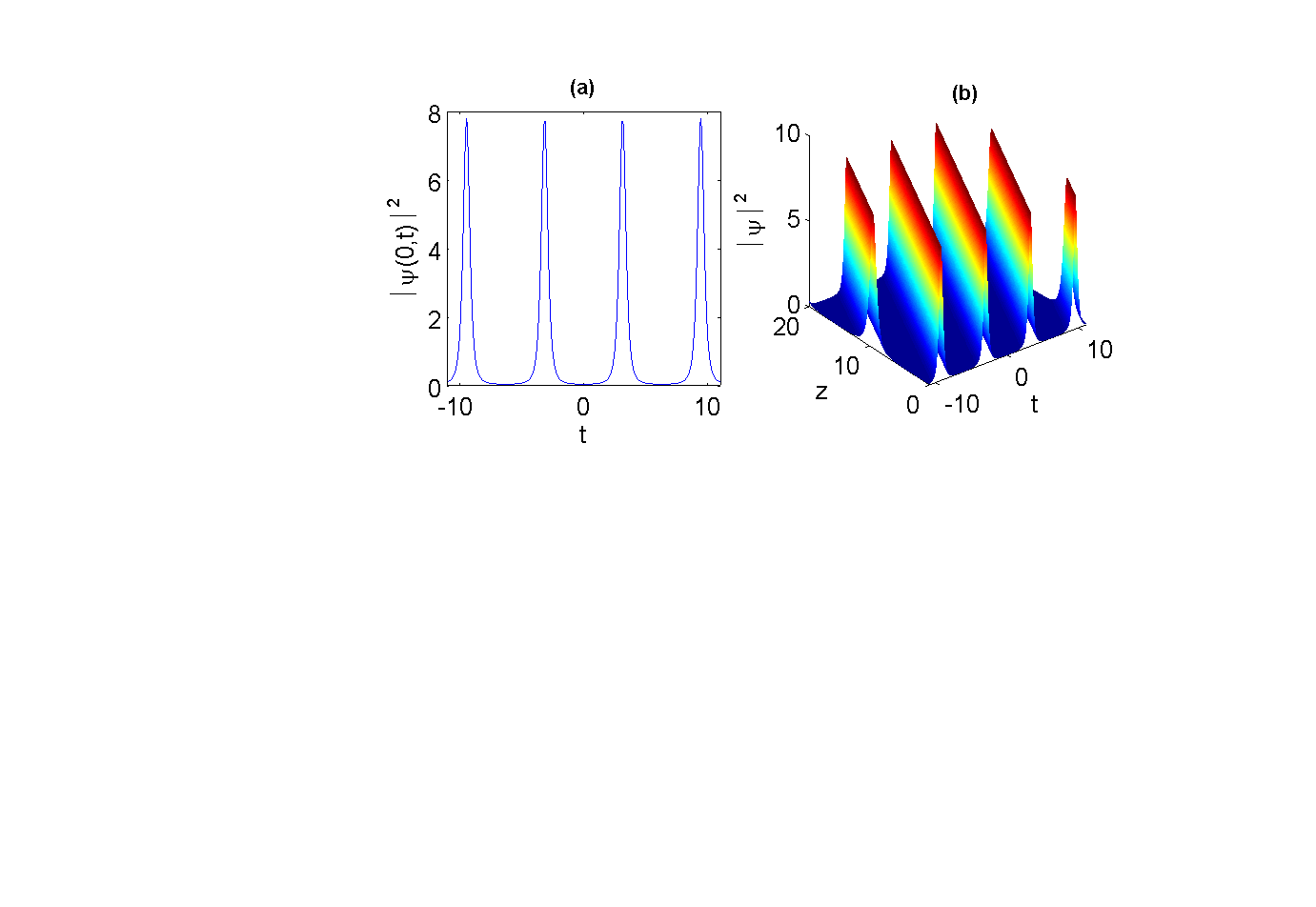} 
\caption{(a) Intensity profile of the chirped periodic wave $\left\vert 
\protect\psi (0,t)\right\vert ^{2}$ as a function of $t$ and its (b)
evolution as computed from Eq. (\ref{33}) for the values $\protect\rho %
=5.0265\times 10^{-10},$ $\protect\kappa =0.15,$ and $\protect\delta =1.5.$
Other parameters are same as given in Fig. 1, and the solitary wave intensity
is normalized by $\left\vert \protect\psi (z,t)\right\vert ^{2}/A^{2}$.}
\label{FIG.2.}
\end{figure}

Figure\ 2(a) depicts the intensity profile of the chirped periodic wave
solution (\ref{33}) for the parameter values $\sigma =-0.1990,$ $\rho
=5.0265\times 10^{-10},$ $\epsilon =3.3511\times 10^{-20},$ $\nu
=-6.2918\times 10^{-14}$, $q=0.5,$ $\kappa =0.15,$ $\eta =0,$ and $\delta
=1.5$, while Fig. 2(b) illustrates its evolution. One observes that the
nonlinear waveform presents an oscillating behavior as it propagates through
the negative index material.

\begin{description}
\item[\textbf{4. Dark solitary waves}] 
\end{description}

We have found a dark solitary wave as 
\begin{equation}
u(x)=\frac{A\tanh (w_{0}(x-\eta ))}{\sqrt{1+B~{\mathrm{sech}}%
^{2}(w_{0}(x-\eta ))}}.  \label{36}
\end{equation}%
In this dark soliton solution, the parameters $A$, $B$ and $w_{0}$ are given
by 
\begin{equation}
A^{2}=-\frac{b}{2c}\pm \frac{1}{2c}\sqrt{b^{2}-4ac},  \label{37}
\end{equation}%
\begin{equation}
B=\frac{2a+2bA^{2}}{4a+bA^{2}},~~~~w_{0}=\sqrt{a+\frac{1}{2}bA^{2}},
\label{38}
\end{equation}%
with $b^{2}-4ac>0$, $A^{2}>0$, $B>-1$ and $a+\frac{1}{2}bA^{2}>0$. Note that
we have the following equation for parameter $A$: $a+bA^{2}+cA^{4}=0$.
Inserting (\ref{36}) into (\ref{2}), we get an exact nonlinearly chirped
dark solitary wave solution for Eq. (\ref{1}) in the form, 
\begin{equation}
\psi (z,t)=\frac{A\tanh (w_{0}(x-\eta ))}{\sqrt{1+B{\mathrm{sech}}%
^{2}(w_{0}(x-\eta ))}}\exp [i\left( \kappa z-\delta t\right) +i\phi (x)].
\label{39}
\end{equation}

The phase function $\phi (x)$ is given by Eq. (\ref{22}) and Eq. (\ref{36})
as%
\begin{eqnarray}
\phi (x) = \frac{3\nu A^{2}C}{4\sigma w_{0}}\ln \left( \frac{C-\tanh
(w_{0}\xi )}{C+\tanh (w_{0}\xi )}\right)  \notag \\
\noalign{\vskip3pt}+\left( \frac{q}{\sigma }+\delta \right) (x-\eta )+\phi
_{0},~~~~~~~~~~~~~~~~~~  \label{40}
\end{eqnarray}%
where the parameter $C$ has the form, 
\begin{equation}
C=\sqrt{1+\frac{1}{B}}=\sqrt{\frac{6a+3bA^{2}}{2a+2bA^{2}}}.  \label{41}
\end{equation}%
It is worth mentioning that the solution (\ref{39}) exists under the
following conditions, 
\begin{equation}
b^{2}-4ac>0,~~~~A^{2}=-\frac{b}{2c}\pm \frac{1}{2c}\sqrt{b^{2}-4ac}>0,
\label{42}
\end{equation}%
\begin{equation}
\frac{2a+2bA^{2}}{4a+bA^{2}}>-1,~~~~a+\frac{1}{2}bA^{2}>0.  \label{43}
\end{equation}%
\begin{figure}[tbp]
\includegraphics[width=\textwidth]{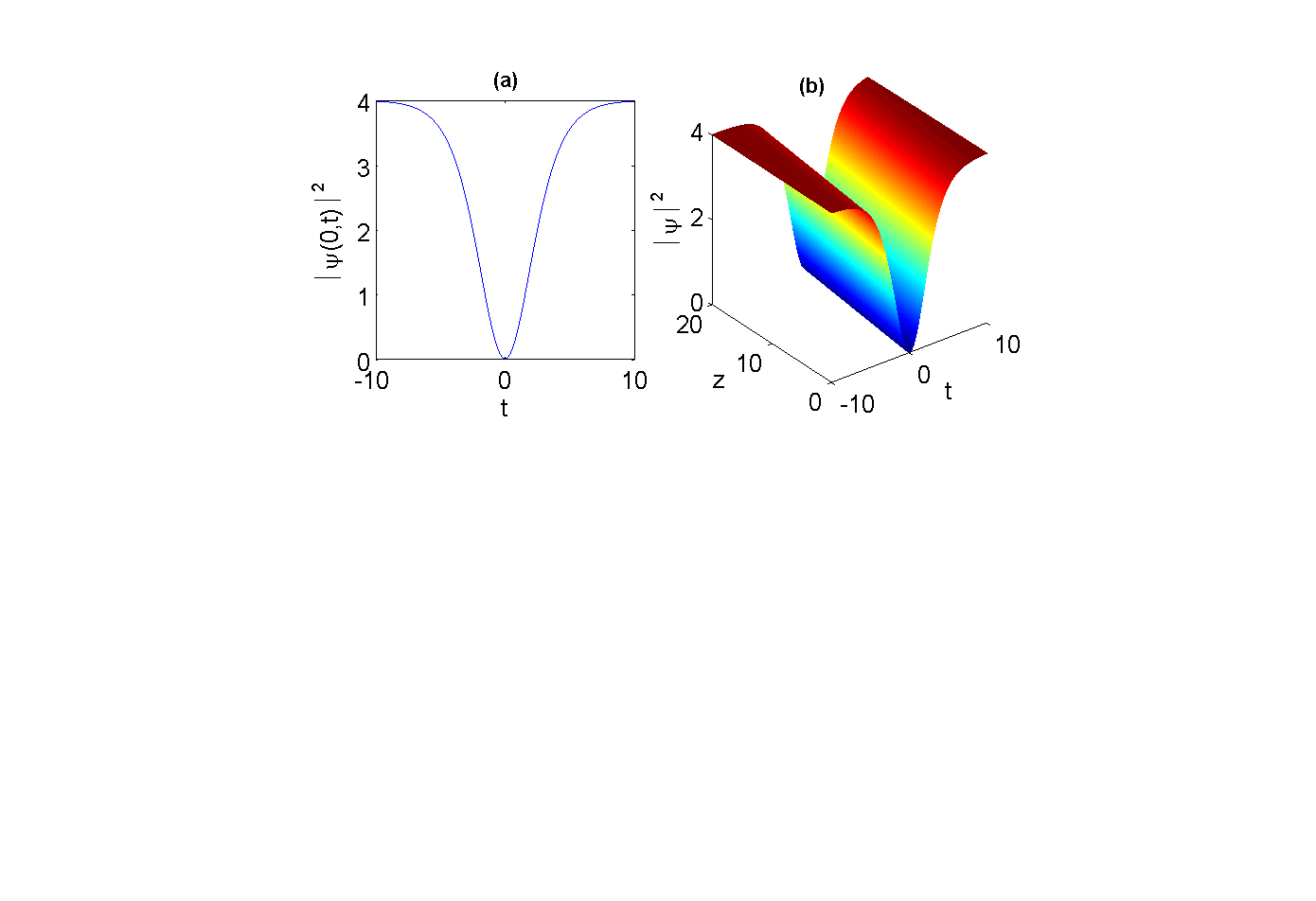} 
\caption{(a) Intensity profile of the chirped dark solitary wave $\left\vert 
\protect\psi (0,t)\right\vert ^{2}$ as a function of $t$ and its (b)
evolution as computed from Eq. (\ref{39}) for the values $\protect\sigma =-0.7954,$
$\protect\rho =1.2566\times 10^{-10},$ $\protect\epsilon =2.095\times
10^{-21},$ $\protect\nu =-4.8195\times 10^{-14},$ $q=0.1,$ $\protect\kappa %
=0.2931,\protect\eta =0,$ and $\protect\delta =2$ . Here the solitary wave
intensity is normalized by $\left\vert \protect\psi (z,t)\right\vert
^{2}/A^{2}$.}
\label{FIG.3.}
\end{figure}
The intensity profile and evolution of the chirped solitary wave solution (%
\ref{39}) in NIMs are shown in Figs. 3(a) and 3(b) respectively. Here we
have taken the parameter values as \cite{Yang}: $\sigma =-0.7954,$ $\rho
=1.2566\times 10^{-10},$ $\epsilon =2.095\times 10^{-21},$ and $\nu
=-4.8195\times 10^{-14}$. Other parameters are $q=0.1,$ $\kappa =0.2931,\eta
=0,$ and $\delta =2$. It can be seen from Fig. 3 that under the influence of
higher-order effects such as pseudo-quintic nonlinearity and self-steepening
effect, the dark-type solitary waves with a nonlinear chirp may exist in
NIMs.

\begin{description}
\item[\textbf{5. Periodic waves connected to dark solitary waves}] 
\end{description}

The dark solitary wave in Eq. (\ref{36}) is connected with the special
periodic solution. We have found such an exact periodic wave solution of Eq.
(\ref{8}) as 
\begin{equation}
u(x)=\frac{A\sin (w(x-\eta ))}{\sqrt{B+\cos ^{2}(w(x-\eta ))}}.  \label{44}
\end{equation}%
In this closed form solution, the parameters $A$, $B$ and $w$ are defined by
the relations, 
\begin{equation}
A^{2}=\frac{b}{2c}\pm \frac{1}{2c}\sqrt{b^{2}-4ac},  \label{45}
\end{equation}%
\begin{equation}
B=\frac{-2a+2bA^{2}}{-4a+bA^{2}},~~~~w=\sqrt{-a+\frac{1}{2}bA^{2}},
\label{46}
\end{equation}%
with $b^{2}-4ac>0$, $A^{2}>0$, $B>0$ and $-a+\frac{1}{2}bA^{2}>0$. Note that
amplitude $A$ satisfies the equation $cA^{4}-bA^{2}+a=0$. Inserting (\ref{44}%
) into (\ref{2}), we get a chirped periodic wave solution for Eq. (\ref{1})
in the form, 
\begin{equation}
\psi (z,t)=\frac{A\sin (w(x-\eta ))}{\sqrt{B+\cos ^{2}(w(x-\eta ))}}\exp
[i\left( \kappa z-\delta t\right) +i\phi (x)].  \label{47}
\end{equation}%
The phase function $\phi (x)$ follows by Eq. (\ref{22}) and Eq. (\ref{44})
as 
\begin{eqnarray}
\phi (x) = \frac{3\nu A^{2}}{2\sigma w}\sqrt{\frac{B+1}{B}}~\mathrm{arctan}%
\left( \sqrt{\frac{B}{B+1}}~\mathrm{tan}\left( w(x-\eta )\right) \right) 
\notag \\
\noalign{\vskip3pt}+\left( \frac{q}{\sigma }+\delta -\frac{3\nu A^{2}}{%
2\sigma }\right) (x-\eta )+\phi _{0}.~~~~~~~~~~~~~~~~~~  \label{48}
\end{eqnarray}%
\begin{figure}[tbp]
\includegraphics[width=\textwidth]{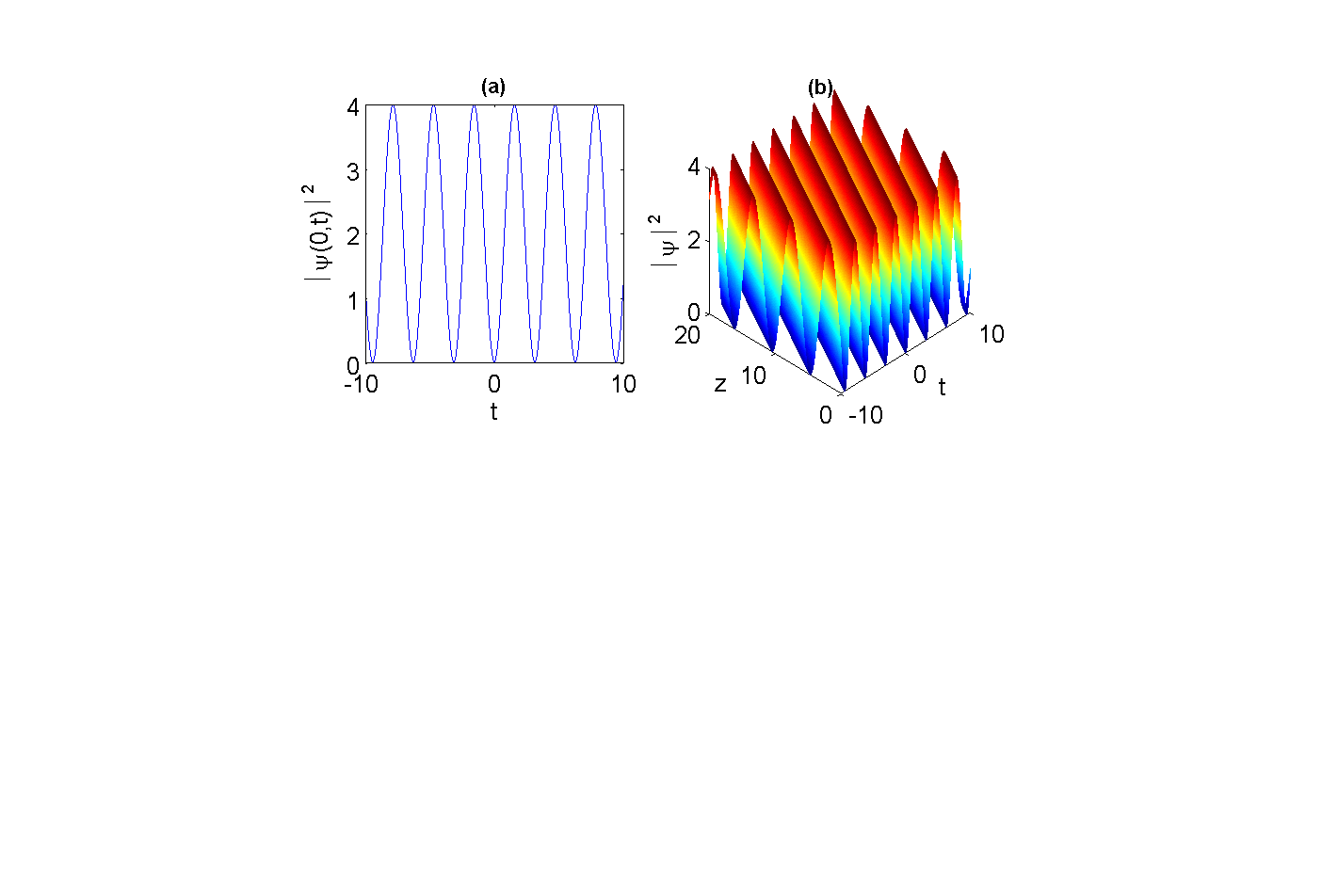} 
\caption{(a) Intensity profile of the chirped periodic wave $\left\vert 
\protect\psi (0,t)\right\vert ^{2}$ as a function of $t$ and its (b)
evolution as computed from Eq. (\ref{47}) for the values $\protect\rho %
=-1.2566\times 10^{-10},$ $\protect\kappa =0.15,$ and $\protect\delta %
=2.5571 $. Other parameters are same as given in Fig. 3, and the solitary
wave intensity is normalized by $\left\vert \protect\psi (z,t)\right\vert
^{2}/A^{2}$. }
\label{FIG.4.}
\end{figure}
An example of the intensity profile and evolution of the chirped periodic
solution (\ref{47}) is shown in Figs. 4(a) and 4(b) respectively, for the
parameter values: $\sigma =-0.7954,$ $\rho =-1.2566\times 10^{-10},$ $%
\epsilon =2.095\times 10^{-21},$ $\nu =-4.8195\times 10^{-14},$ $q=0.1,$ $%
\kappa =0.15,\eta =0,$\ and $\delta =2.5571$.

\begin{description}
\item[\textbf{6. Gray solitary waves}] 
\end{description}

We have also found an exact solution for Eq. (\ref{18}) of the form, 
\begin{equation}
u(x)=\frac{A}{\sqrt{1+B~{\tanh }^{2}(w_{0}(x-\eta ))}},  \label{49}
\end{equation}%
where 
\begin{equation}
B=\lambda -3\pm \sqrt{\lambda (\lambda -3)},~~~~\lambda =\frac{3b^{2}}{4ac},
\label{50}
\end{equation}

\begin{equation}
A^{2}=-\frac{3b(B+1)}{2c(B+3)},~~~~w_{0}=\sqrt{\frac{aB}{3+2B}},  \label{51}
\end{equation}

\noindent with the requirement as $B>-1$, $\lambda (\lambda -3)>0$, $bc<0$
and $aB>0$. Note that the parameter $B$ satisfies the equation $%
(B+3)^{2}=\lambda (2B+3)$. Thus in the general case, we have $B>-1$ and we
obtain a gray solitary wave solution (i.e., a dark pulse with a nonzero
minimum in intensity) as 
\begin{equation}
\psi (z,t)=\frac{A}{\sqrt{1+B~{\tanh }^{2}(w_{0}(x-\eta ))}}\exp [i(\kappa
z-\delta t)+i\phi (x)].  \label{52}
\end{equation}%
The phase $\phi (x)$ given in Eq. (\ref{52}) for the case $-1<B<0$ is 
\begin{eqnarray}
\phi (x) = \frac{3\nu A^{2}\sqrt{-B}}{4\sigma w_{0}(1+B)}\ln \left( \frac{1-%
\sqrt{-B}\tanh (w_{0}(x-\eta ))}{1+\sqrt{-B}\tanh (w_{0}(x-\eta ))}\right) 
\notag \\
\noalign{\vskip3pt}+\left( \delta +\frac{q}{\sigma }+\frac{3\nu A^{2}}{%
2\sigma (1+B)}\right) (x-\eta )+\phi _{0}.~~~~~~~~~~~~~~  \label{53}
\end{eqnarray}%
\begin{figure}[tbp]
\includegraphics[width=\textwidth]{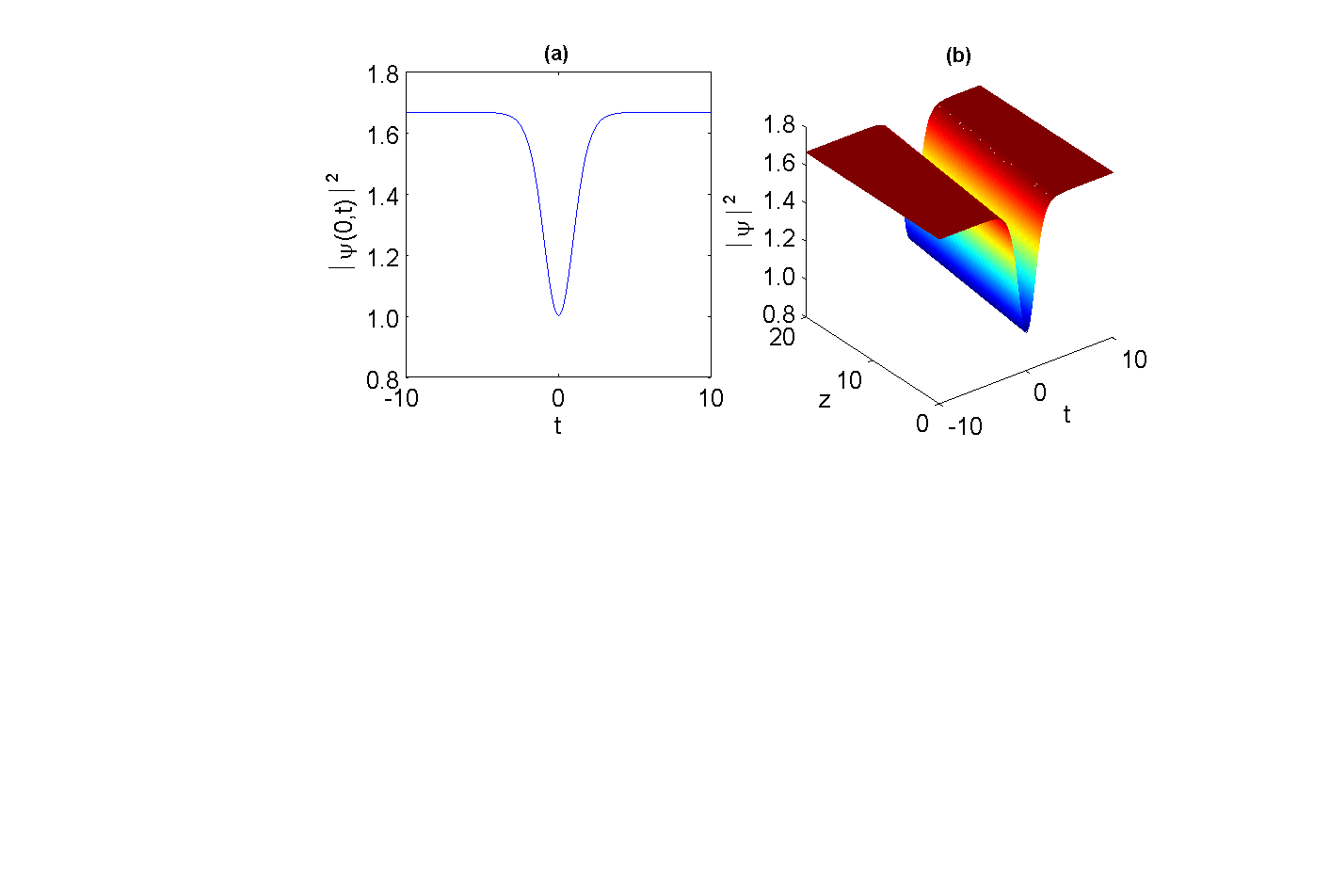} 
\caption{(a) Intensity profile of the chirped gray solitary wave $\left\vert 
\protect\psi (0,t)\right\vert ^{2}$ as a function of $t$ and its (b)
evolution as computed from Eq. (\ref{52}) for the values $\protect\rho %
=-1.2566\times 10^{-10},$ $\protect\epsilon =-2.095\times 10^{-21},$ $%
\protect\kappa =0.15,$\ $B=-0.4,$ and $\protect\delta =20.0581.$ Other
parameters are same as given in Fig. 3, and the solitary wave intensity is
normalized by $\left\vert \protect\psi (z,t)\right\vert ^{2}/A^{2}$.}
\label{FIG.5.}
\end{figure}
Figures 5(a) and 5(b) illustrate the intensity profile and evolution of the
chirped solitary wave solution (\ref{52}). Here the used parameter values: $%
\sigma =-0.7954,$ $\rho =-1.2566\times 10^{-10},$ $\epsilon =-2.095\times
10^{-21},$ $\nu =-4.8195\times 10^{-14}$, $q=0.1,$ $\kappa =0.15,$\ $B=-0.4,$
$\eta =0,$ and $\delta =20.0581$.

\begin{description}
\item[\textbf{7. Periodic waves connected to gray solitary waves}] 
\end{description}

The gray solitary wave in Eq. (\ref{49}) is connected with the special
periodic solution. We have found such an exact periodic wave solution of Eq.
(\ref{8}) in the form,%
\begin{equation}
u(x)=\frac{A}{\sqrt{1+B~\tan ^{2}(w(x-\eta ))}},  \label{54}
\end{equation}%
where 
\begin{equation}
B=3-\lambda \pm \sqrt{\lambda (\lambda -3)},~~~~\lambda =\frac{3b^{2}}{4ac},
\label{55}
\end{equation}
\begin{figure}[tbp]
\includegraphics[width=\textwidth]{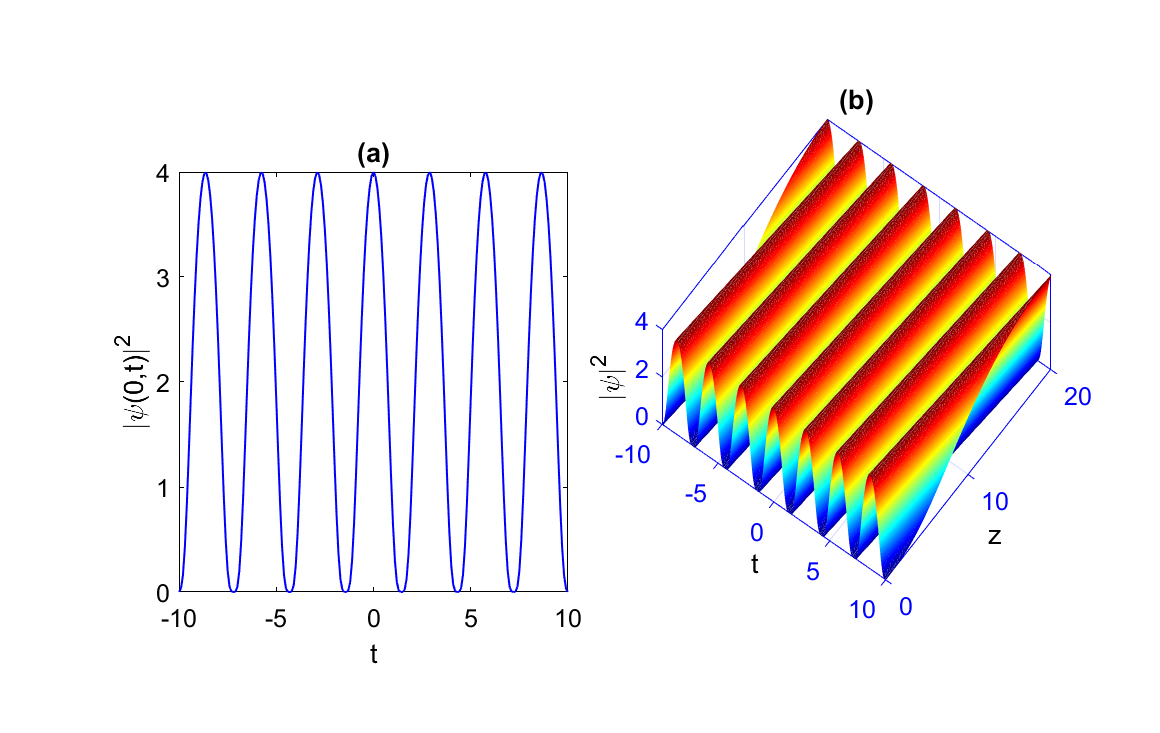} 
\caption{(a) Intensity profile of the chirped periodic wave $\left\vert 
\protect\psi (0,t)\right\vert ^{2}$ as a function of $t$ and its (b)
evolution as computed from Eq. (\ref{57}) for the values $\protect\kappa=1.865,$ $B=0.4,$ and $\protect\delta =0.314.$ Other parameters are same as given in Fig. 3, and the solitary wave intensity is normalized by $\left\vert \protect\psi (z,t)\right\vert ^{2}/A^{2}$.}
\label{FIG.6.}
\end{figure}
\begin{equation}
A^{2}=-\frac{3b(1-B)}{2c(3-B)},~~~~w=\sqrt{\frac{aB}{3-2B}},  \label{56}
\end{equation}

\noindent with the requirement $B>0$, $\lambda(\lambda-3) > 0$, $%
bc(1-B)(3-B)<0$ and $a(3-2B)>0$. Note that the parameter $B$ satisfies the
equation $(3-B)^{2}=\lambda(3-2B)$.

Thus we obtain a chirped periodic wave solution of Eq. (\ref{1}) as%
\begin{equation}
\psi (z,t)=\frac{A}{\sqrt{1+B~{\tan }^{2}(w(x-\eta ))}}\exp [i(\kappa
z-\delta t)+i\phi (x)].  \label{57}
\end{equation}%
The phase $\phi (x)$ follows by Eq. (\ref{22}) and Eq. (\ref{54}) as%
\begin{eqnarray}
\phi (x) = \frac{3\nu A^{2}\sqrt{B}}{2\sigma w\left( B-1\right) }~\mathrm{%
arctan}\left[ \sqrt{B}~\mathrm{tan}\left( w(x-\eta )\right) \right]  \notag
\\
\noalign{\vskip3pt}+\left( \frac{q}{\sigma }+\delta +\frac{3\nu A^{2}}{%
2\sigma \left( 1-B\right) }\right) (x-\eta )+\phi _{0}.~~~~~~~~~~~~~~~
\label{58}
\end{eqnarray}%
The profile and evolution of the chirped periodic wave solution (\ref{57})
are shown in Figs. 6(a) and 6(b), respectively. Here we have taken the \
parameter values \cite{Yang}: $\sigma =-0.7954,$ $\rho =1.2566\times
10^{-10},$ $\epsilon =2.095\times 10^{-21},$ and $\nu =-4.8195\times
10^{-14} $. Other parameters are $q=0.1,$ $\kappa =1.865,$\ $B=0.4,$ $\eta
=0,$ and $\delta =0.314$.

\begin{description}
\item[\textbf{8. Kink-type solitary waves}] 
\end{description}

The solution of Eq. (\ref{18}), in particular case when $3b^{2}/16ac=1$, can be written in the form,
\begin{equation}
u(x)=A\sqrt{1\pm {\tanh }(w_{0}(x-\eta ))},  \label{59}
\end{equation}%
where%
\begin{equation}
A^{2}=-\frac{2a}{b},~~~~w_{0}=\sqrt{-a},  \label{60}
\end{equation}

\noindent with conditions $a<0$ and $b>0$. Thus the chirped kink-type
solitary wave solution of Eq. (\ref{1}) is 
\begin{equation}
\psi (z,t)=A\sqrt{1\pm {\tanh }(w_{0}(x-\eta ))}\exp [i(\kappa z-\delta
t)+i\phi (x)].  \label{61}
\end{equation}%
In this case, the phase function $\phi (x)$ is given by Eq. (\ref{22}) and
Eq. (\ref{59}) as 
\begin{eqnarray}
\phi (x) = \pm \frac{3\nu A^{2}}{2\sigma w_{0}}\ln [\cosh (w_{0}(x-\eta ))]
\notag \\
\noalign{\vskip3pt}+\left( \delta +\frac{q}{\sigma }+\frac{3\nu A^{2}}{%
2\sigma }\right) (x-\eta )+\phi _{0}.  \label{62}
\end{eqnarray}%
Figures 7(a) and 7(b) depict the evolution of the intensity wave profile of
kink-shaped (upper sign) and anti-kink-shaped (lower sign) solitary waves as
computed from Eq. (\ref{61}) for the same parameter values $\sigma =-0.7954,$
$\rho =-1.2566\times 10^{-10},$ $\epsilon =-2.095\times 10^{-21},$ and $\nu
=-4.8195\times 10^{-14}$. Other parameters are $q=0.1,$ $\kappa =1.2,\eta
=0, $ and $\delta =26.1937$.\newline
\begin{figure}[tbp]
\includegraphics[width=\textwidth]{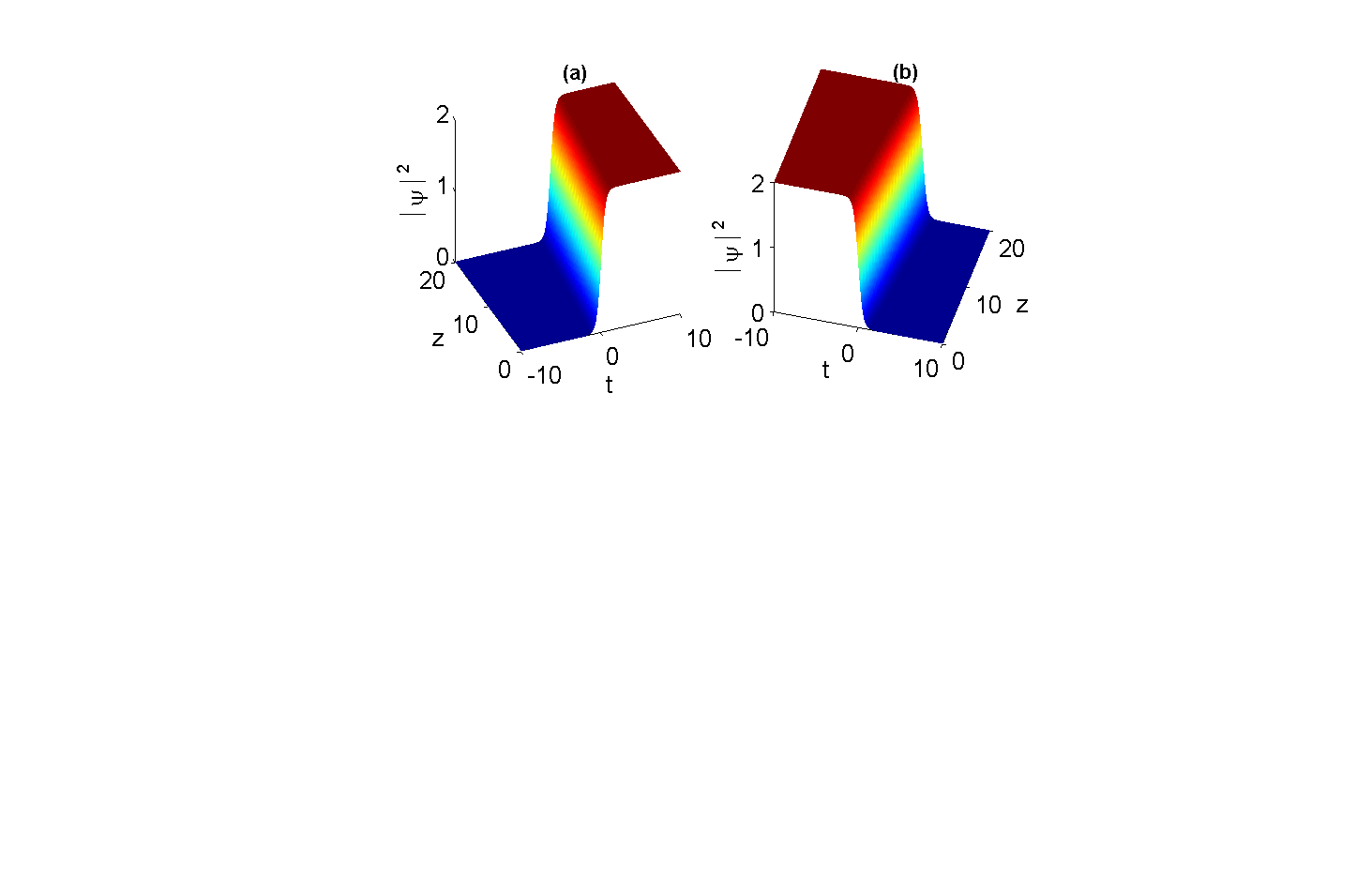} 
\caption{(a)\ Intensity profile of the chirped kink-type solitary wave $\left\vert 
\protect\psi (0,t)\right\vert ^{2}$ as a function of $t$ and its (b)
evolution as computed from Eq. (\ref{61}) for the values $\protect\rho %
=-1.2566\times 10^{-10},$ $\protect\epsilon =-2.095\times 10^{-21},$ $%
\protect\kappa =1.2,$ and $\protect\delta =26.1937.$ Other parameters are
same as given in Fig. 3, and the solitary wave intensity is normalized by $%
\left\vert \protect\psi (z,t)\right\vert ^{2}/A^{2}$.}
\label{FIG.7.}
\end{figure}

\section{Stability analysis of periodic and solitary waves}

We consider in this section the analytical stability analysis of soliton
solutions and periodic waves for the generalized NLSE (\ref{1}). Our
approach is based on the theory of nonlinear dispersive waves in nonlinear
optics and fluid dynamics \cite{Vladimir,KrTr,TK,Whit}. For our purpose, we
develop the dynamics of dispersive waves in the following form, 
\begin{equation}
\psi (z,t)=F(\Delta \omega )\exp [i\Theta (z,t)],  \label{63}
\end{equation}%
where the amplitude $F(\Delta \omega )$ and phase $\Theta (z,t)$ are real
functions, and $\Delta \omega (z,t)$ is the frequency shift of nonlinear
dispersive wave. It is worth noting that $\psi (z,t)$ is slowly varying
amplitude for the pulse envelope. Hence the frequency of nonlinear
dispersive wave is $\omega _{0}+\Delta \omega $ where $\omega _{0}$ is the
carrier frequency. However, we use in this section more simple notation $%
\Delta \omega (z,t)\equiv \omega (z,t)$ for the frequency shift of the
dispersive waves. Hence, the amplitude in Eq. (\ref{63}) is $F(\omega )$,
and the wave number shift $k$ and frequency shift $\omega $ of nonlinear
dispersive waves are 
\begin{equation}
k(\omega )=\frac{\partial \Theta }{\partial z},~~~~\omega =-\frac{\partial
\Theta }{\partial t}.  \label{64}
\end{equation}%
We have by definition that $\Theta _{zt}=k_{t}$ and $\Theta _{tz}=-\omega
_{z}$ which yield the equation, 
\begin{equation}
\frac{\partial \omega }{\partial z}+\frac{\partial k(\omega )}{\partial t}=0.
\label{65}
\end{equation}%
We assume in our approach that $\omega (z,t)=\tilde{\omega}(Z,T)$ and $%
k(z,t)=\tilde{k}(Z,T)$ where $Z=\varepsilon z$ and $T=\varepsilon t$ with $%
\varepsilon \ll 1$ are slow variables. Here $\tilde{\omega}(Z,T)$ and $%
\tilde{k}(Z,T)$ are slow varying functions of variables $Z$ and $T$.

Thus this approach to stability analysis of solitons and periodic waves \cite%
{Vladimir,KrTr,TK} is based on the method of slow variables. Note that the
generalized NLSE (\ref{1}) is derived in quasi-monochromatic approximation
assuming that the parameter $\varepsilon\equiv\Delta\omega_{s}/\omega_{0}$
is small ($\varepsilon\ll 1$). Here $\Delta\omega_{s}$ and $\omega_{0}$ are
the spectral width of the pulse and the carrier frequency respectively.
Hence the condition $\varepsilon\ll 1$ is satisfied for the generalized NLSE
(\ref{1}). The substitution of Eq. (\ref{63}) (with $\Delta\omega\equiv%
\omega $) into the generalized NLSE (\ref{1}) leads to series of nonlinear
equations. The equation in zero order to small parameter $\varepsilon$ is 
\begin{equation}
k(\omega )=k_{0}(\omega )+\Gamma (\omega )F^{2}(\omega )-\epsilon
F^{4}(\omega ),  \label{66}
\end{equation}%
where $k_{0}(\omega )$ and $\Gamma (\omega )$ are 
\begin{equation}
k_{0}(\omega )=-\frac{\sigma}{2} \omega ^{2},~~~~\Gamma (\omega )=\rho -\nu
\omega .  \label{67}
\end{equation}%
The parameter $\Gamma (\omega )$ is nonlinear coefficient renormalized by
self-steepening effect. Equations (\ref{1}), (\ref{63}), (\ref{64}) in the
first order to small parameter $\varepsilon $ yield 
\begin{equation}
\frac{\partial F}{\partial z}+k_{0}^{^{\prime }}(\omega )\frac{\partial F}{%
\partial t }=-\frac{1}{2}k_{0}^{^{\prime \prime }}(\omega )F\frac{\partial
\omega }{\partial t }+3\nu F^{2}\frac{\partial F}{\partial t },  \label{68}
\end{equation}%
with $k_{0}^{^{\prime }}(\omega )=dk_{0}(\omega )/d\omega =-\sigma\omega$.
Thus Eq. (\ref{66}) is the nonlinear dispersion relation and Eq. (\ref{68})
is the equation for amplitude $F(\omega )$ of nonlinear waves. Note that Eq.
(\ref{65}) for varying frequency shift $\omega (z,t )$ can also be written
in the form, 
\begin{equation}
\frac{\partial \omega }{\partial z}+k^{^{\prime }}(\omega )\frac{\partial
\omega }{\partial t}=0,  \label{69}
\end{equation}%
where the function $k^{^{\prime }}(\omega )$ is 
\begin{equation}
k^{^{\prime }}(\omega )=k_{0}^{^{\prime }}(\omega )-\nu F^{2}(\omega
)+2\Gamma (\omega)F(\omega )F^{^{\prime }}(\omega )-4\epsilon F^{3}(\omega
)F^{^{\prime }}(\omega ),  \label{70}
\end{equation}%
with $F^{^{\prime}}(\omega )=dF(\omega )/d\omega $. Equation (\ref{70})
allows us to write Eq. (\ref{68}) in the following form, 
\begin{eqnarray}
\frac{\partial F}{\partial z}+k^{^{\prime }}(\omega )\frac{\partial F}{%
\partial t }=-\frac{1}{2}k_{0}^{^{\prime \prime }}(\omega )F\frac{\partial
\omega }{\partial t }+2\nu F^{2}\frac{\partial F}{\partial t }  \notag \\
\noalign{\vskip3pt} +2(\Gamma (\omega)-2\epsilon F^{2})FF^{^{\prime }}\frac{%
\partial F}{\partial t },~~~~~~~~~~~~~~  \label{71}
\end{eqnarray}
with $k_{0}^{^{\prime \prime }}(\omega )=-\sigma$. The system of Eqs. (\ref%
{69}) and (\ref{71}) based on the method of slow variables can be hyperbolic
or elliptic. We first consider the case when this system of equations is
hyperbolic. The characteristics connected to the hyperbolic system of Eqs. (%
\ref{69}) and (\ref{71}) are given by 
\begin{equation}
\frac{dt }{dz}=k^{^{\prime }}(\omega ),~~~~\frac{d\omega }{dz}=0,  \label{72}
\end{equation}%
\begin{equation}
\frac{dF}{dz}=\left(\frac{1}{2}\sigma+2\nu FF^{^{\prime }}+ 2(\Gamma
(\omega)-2\epsilon F^{2})(F^{^{\prime }})^{2}\right)F\frac{\partial \omega }{%
\partial t } .  \label{73}
\end{equation}%
It follows from Eq. (\ref{72}) that $dF/dz=F^{^{\prime }}(\omega )d\omega
/dz=0$. Hence Eq. (\ref{73}) yields the equation, 
\begin{equation}
(\Gamma (\omega)-2\epsilon F^{2})(F^{^{\prime }})^{2}+\nu FF^{^{\prime }}+%
\frac{\sigma}{4}=0 .  \label{74}
\end{equation}%
Thus we have the following nonlinear differential equation for the function $%
F(\omega )$, 
\begin{equation}
\frac{dF(\omega )}{d\omega }=-\frac{\nu F(\omega )}{2G(\omega )}\pm \frac{%
\sqrt{\nu ^{2}F^{2}(\omega )-\sigma G(\omega)}}{2G(\omega )},  \label{75}
\end{equation}%
where $G(\omega)=\Gamma (\omega)-2\epsilon F^{2}(\omega)$. Equations (\ref%
{70}), (\ref{72}), and (\ref{75}) yield the characteristic equation, 
\begin{eqnarray}
\frac{dt }{dz}=-\sigma\omega -2\nu F^{2}(\omega )~~~~~~~~~~~~~~  \notag \\
\noalign{\vskip3pt} \pm F(\omega )\sqrt{\nu ^{2}F^{2}(\omega )-\sigma(\Gamma
(\omega)-2\epsilon F^{2}(\omega))}.  \label{76}
\end{eqnarray}
It follows from this equation that the system of Eqs. (\ref{69}) and (\ref%
{71}) with infinitesimal amplitude $F(\omega )$ is hyperbolic when the
condition $\sigma\Gamma (\omega )< 0$ is satisfied, and the system of
equations with infinitesimal amplitude $F(\omega )$ is elliptic for the
following condition $\sigma\Gamma (\omega )> 0$ or in an explicit form $%
\sigma\rho>\sigma\nu\omega$.

Equations (\ref{75}) and (\ref{76}) lead to physical interpretation of
stability for the solitons and periodic solutions of the extended NLSE (\ref%
{1}). We may assert \cite{KrTr,TK} that the soliton is stable when it can
not radiate the nonlinear dispersive waves. Note that the outgoing nonlinear
dispersive waves with infinitesimal amplitude $F(\omega )$, connected to
such radiation process, exist only in the case when above system of
equations is hyperbolic. Moreover, in the case of elliptic equations the
problem of optical pulse radiation is not correct from the mathematical
point of view. For the infinitesimal amplitude $F(\omega )$ this takes place
in the case when $\sigma\Gamma (\omega )> 0$. This relation follows from
Eqs. (\ref{75}) and (\ref{76}) because the outgoing nonlinear dispersive
waves not exist when the square root in these equations is imaginary and
hence the system of Eqs. (\ref{69}) and (\ref{71}) is elliptic. Thus this
physical interpretation of stability of soliton solution yields the
stability domain $\mathcal{D}_{st}$ given by inequality $\sigma\rho>\sigma%
\nu\omega$.

Note that the function $\Theta(z,t)$ is given by Eqs. (\ref{2}) and (\ref{63}%
) as $\Theta(z,t)=\kappa z-\delta t+\phi (x)$. Hence the frequency shift $%
\omega=-\partial\Theta/\partial t$ is given by $\omega = \delta
-\partial\phi /\partial t$, which leads by Eq. (\ref{6}) to the following
explicit form, 
\begin{equation}
\omega(x) = -\frac{q}{\sigma}-\frac{3\nu}{2\sigma}u^{2}(x),  \label{77}
\end{equation}%
where the local frequency shift $\omega(x) $ is the function of variables $z$
and $t$ ($x=t-qz$). We emphasize that Eqs. (\ref{77}) and (\ref{7}) coincide
with our notation as $\Delta\omega\equiv\omega$. However, Eq. (\ref{77}) has
different meaning because it describes the relation of the local frequency
shift $\omega(x)$ of nonlinear dispersive waves from the intensity $%
I(x)=u^{2}(x)$ of the periodic or solitary waves propagating in negative
index materials. Thus we assume that the chirp of traveling-wave governing
by generalized NLSE (\ref{1}) is equal to the local frequency shift $%
\omega(x)$ of excited infinitesimal nonlinear dispersive waves.

It follows from Eq. (\ref{77}) that the found stability condition $%
\sigma\rho>\sigma\nu\omega$ yields for the generalized NLSE (\ref{1}) the
following stability criterion, 
\begin{equation}
u^{2}(x)> -\frac{2(\sigma\rho+q\nu)}{3\nu^{2}},  \label{78}
\end{equation}%
where the variable $x$ belongs to the interval $-\infty<x<+\infty$. The
stability criterion in Eq. (\ref{78}) is equivalent to the following
condition, 
\begin{equation}
I_{m}+\frac{2}{3\nu^{2}}(\sigma\rho+q\nu)>0,  \label{79}
\end{equation}%
where $I_{m}\equiv\mathrm{min}~ (u^{2}(x))$ for $-\infty<x<+\infty$. In the
case when intensity $I(x)=u^{2}(x)$ of the solitary or periodic wave is zero
for some value of $x=t-qz$ or tends to zero for $x\rightarrow \pm\infty$ the
parameter $I_{m}$ in Eq. (\ref{79}) is $I_{m}=0$. In this important case the
stability condition in Eq. (\ref{79}) has the form, 
\begin{equation}
\sigma\rho+q\nu>0.  \label{80}
\end{equation}
\begin{figure}[tbp]
\includegraphics[width=\textwidth]{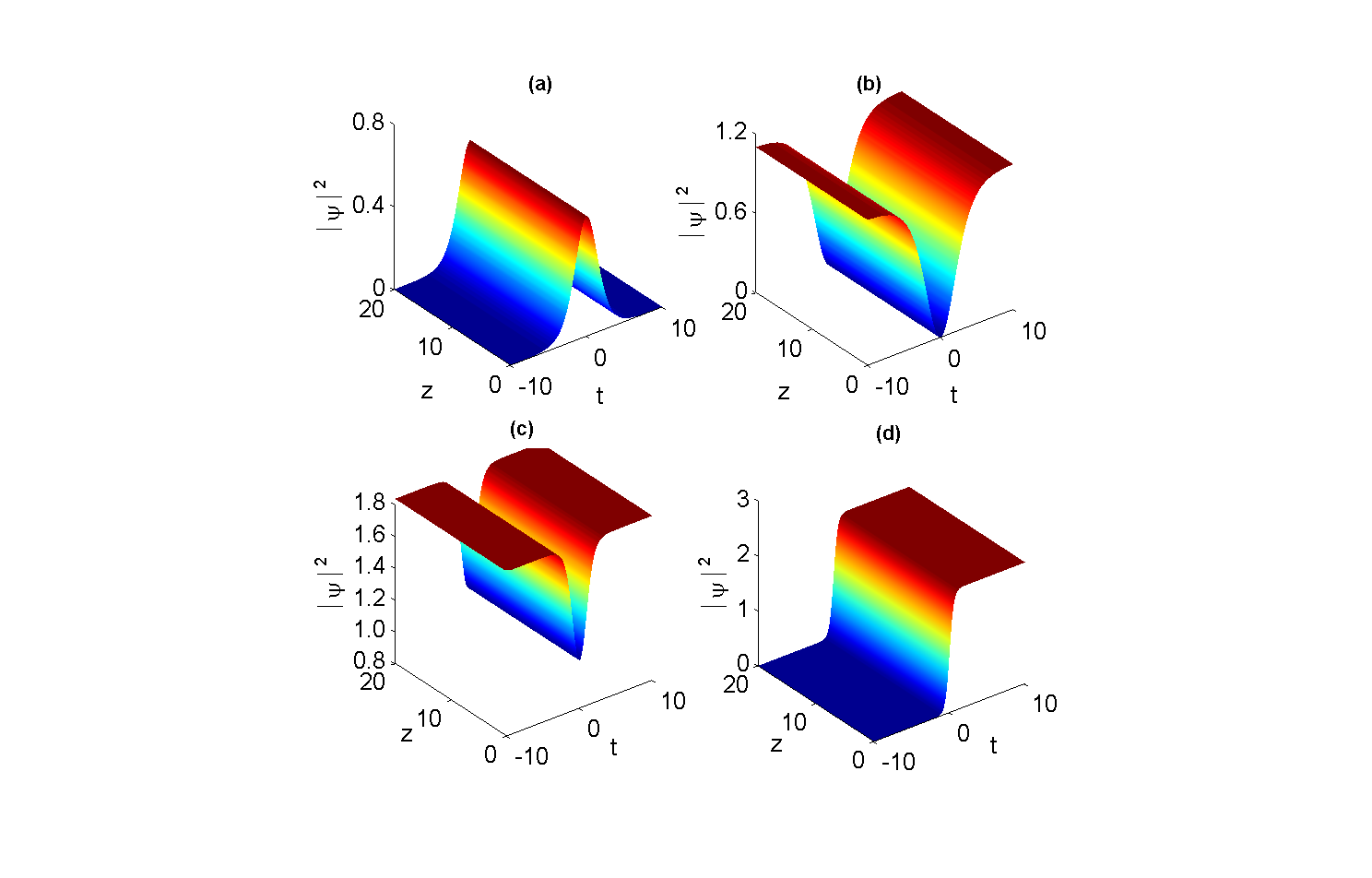} 
\caption{ The numerical evolution of an initial (a) bright, (b) dark, (c) gray, and (d) kink pulses whose amplitude $10\%$ smaller than the exact solution. The parameters are the same as in Figs. 1(b), 3(b),
5(b), and 7(b) respectively. }
\label{FIG.8.}
\end{figure}
We emphasize that the criterion in Eq. (\ref{80}) does not depend on
pseudo-quintic nonlinearity coefficient $\epsilon$ because this stability
condition is the particular case of the general criterion in Eq. (\ref{79})
when $I_{m}=0$. However, the general criterion in Eq. (\ref{79}) depends on
the parameter $\epsilon$ because the intensity $I(x)=u^{2}(x)$ of
propagating waves and hence the parameter $I_{m}$ are the functions of
coefficient $\epsilon$ in the case when $I_{m}\neq 0$.

Note that the stability criterion in Eq. (\ref{80}) is equivalent to the
condition $b>0$ where the parameter $b$ is introduced in Eq. (\ref{9}). For
an example, we have $I_{m}=0$ for the bright soliton given in Eq. (\ref{30}%
), and hence the stability condition for this soliton solution is $b>0$.
Thus the necessary conditions ($a<0$, $b>0$, and $3b^{2}>16ac$) for
existence of this soliton are sufficient in appropriate experimental
situation. We have also the parameter $I_{m}=0$ for the dark soliton
solution in Eq. (\ref{39}), and hence in this case the stability condition
is $b>0 $. However for the gray soliton solution in Eq. (\ref{52}) we have
two cases: (1) $I_{m}=A^{2}/(1+B)$ for $B>0$, and (2) $I_{m}=A^{2}$ for $%
-1<B<0$. Hence the stability condition for gray soliton in these two cases
is given by Eq. (\ref{79}) with appropriate values for parameter $I_{m}$. In
the case of kink-type wave given in Eq. (\ref{61}) we have $I_{m}=0$. Hence
the conditions $a<0$ and $b>0$ with the constraint $3b^{2}/16ac=1$ are
necessary and sufficient for existence of this kink-type waves in
appropriate experimental situation. In the case of periodic waves presented
in sec. III the stability criteria given in Eqs. (\ref{79}) and (\ref{80})
are also applicable.

\section{Numerical analysis}

\begin{figure}[tbp]
\includegraphics[width=\textwidth]{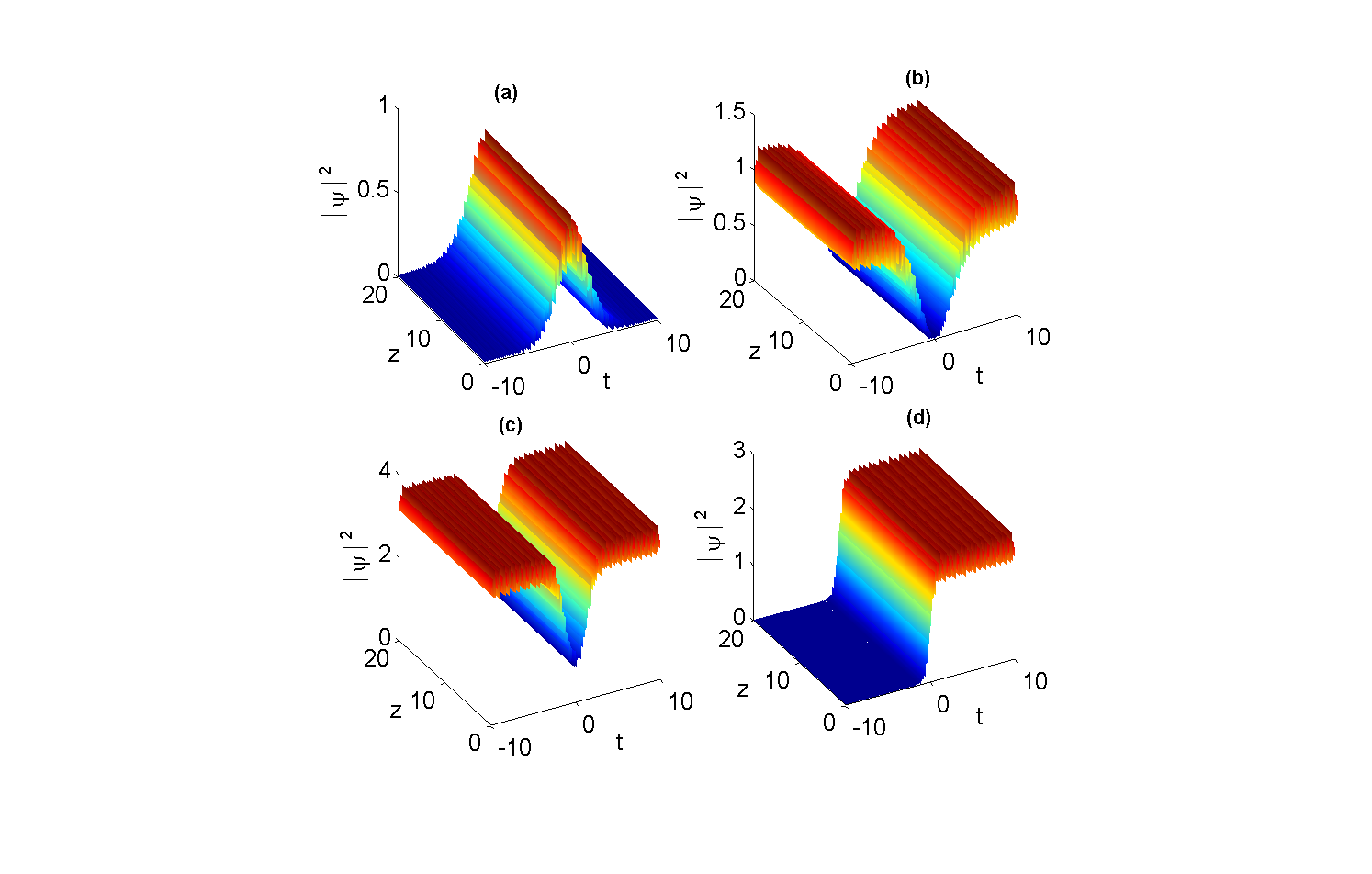} 
\caption{ The numerical evolution of (a) the chirped bright solitary wave
solution (\ref{30}), (b) the chirped dark solitary wave solution (\ref{39}), (c) the chirped gray solitary wave solution (\ref{52}), and (d) the chirped kink solitary wave solution (\ref{61}) under the perturbation of white noise whose maximal value is $0.1$. The parameters are the same as in Figs. 1(b), 3(b), 5(b), and 7(b) respectively. }
\label{FIG.9.}
\end{figure}
\begin{figure}[tbp]
\includegraphics[width=\textwidth]{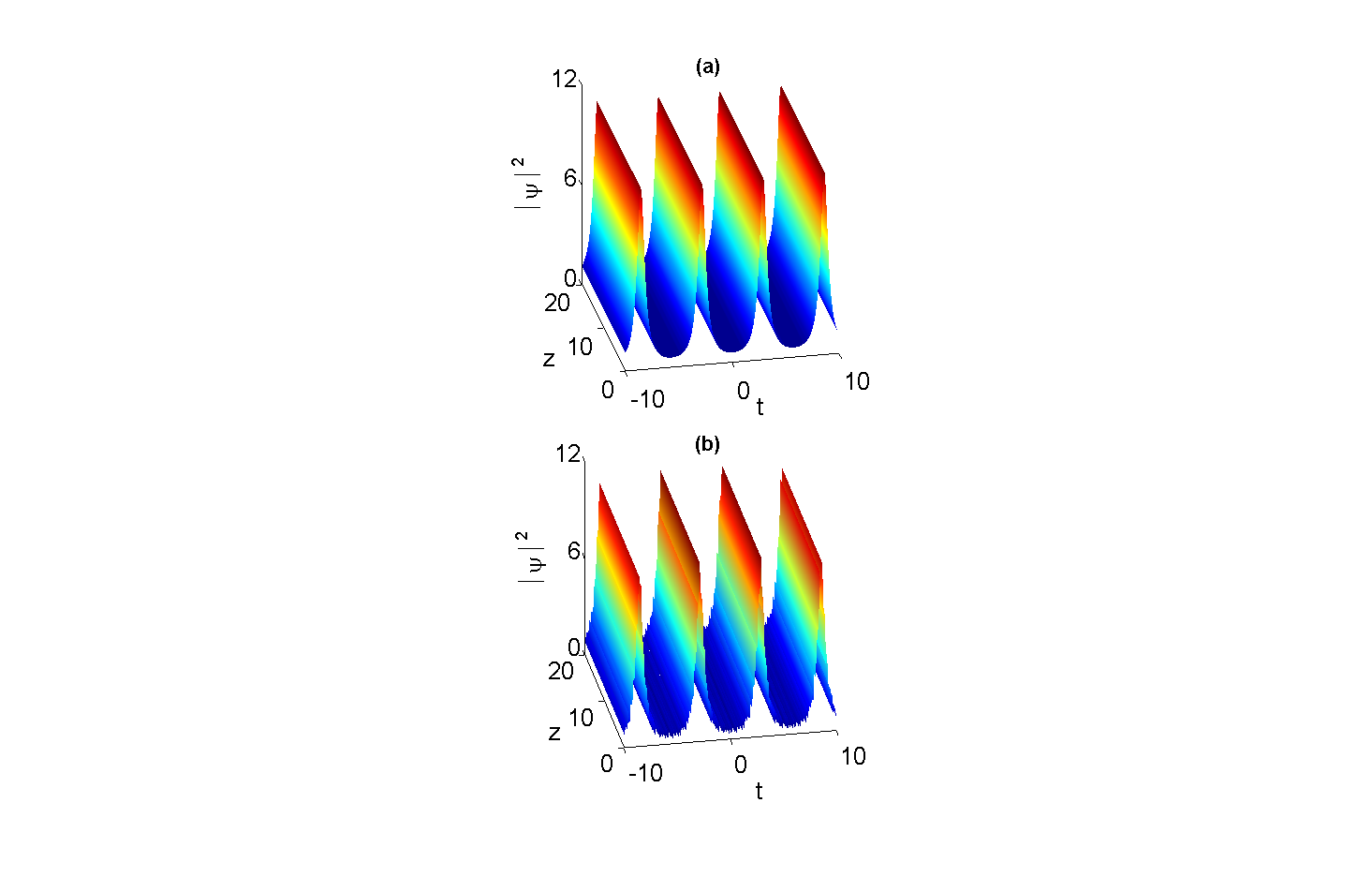} 
\caption{ The numerical evolution of (a) the chirped periodic solution (\ref{33}) with an initial wave amplitude $10\%$ smaller than the exact solution; (b) the chirped periodic solution (\ref{33}) under perturbation of white noise whose maximal value is $0.1$. The parameters are the same as in Fig. 2.}
\label{FIG.10.}
\end{figure}

In this section, we investigate the stability of the obtained periodic and
solitary waves by using directly numerical simulations. It should be
mentioned that only stable (or weakly unstable) solitary waves are promising
for experimental observations and practical applications. Therefore,\ it is
worthy to study the stability of these nonlinearly chirped structures with
respect to finite perturbations. One notes that bright solitons are found to
be stable in focusing Kerr-type media \cite{Aitchison}. Moreover, important
results have established the essential role played by higher-order
nonlinearities in stabilizing the propagation of solitons. In this context,
it was shown that the competition between focusing third-order and
defocusing fifth-order nonlinearities supports stable soliton solutions \cite%
{Skarka}. For our case of study, the generalized NLSE\ (\ref{1}) involves
such cubic-quintic nonlinearites, in addition to a derivative Kerr
nonlinearity. It is therefore not unreasonable to conjecture that the
obtained solutions are stable. However, a detailed analysis is required in
order to strictly answer the stability problem of such privileged chirped
nonlinear waveforms.

In order to analyze the stability of these chirped structures, we performed
direct numerical simulations of Eq. (\ref{1}) initialized with our computed
solutions with added white noise and amplitude perturbation \cite{He,R2}. We
first perturb the amplitude ($10\%$) in the initial distribution. Figures
8(a)-8(d) show the evolution plots of chirped solitary-wave solutions (\ref%
{30}), (\ref{39}), (\ref{52}), and (\ref{61}) in which the amplitude in the
initial distributions is perturbed. From these figures, one observes
remarkably stable propagation of the localized solutions under the amplitude
perturbation. Second, we add white noise ($10\%$) in the initial pulses. The
numerical results are shown in Figs. 9(a)-9(d) in which the initial pulses
are perturbed by white noise. Compared with Figs. 1(b), 3(b), 5(b), and
7(b), one can see that despite adding a white noise perturbation, the
solutions remains intact after propagating a distance of twenty dispersion
lengths. Thus, we can conclude that the obtained chirped bright, dark, gray,
and kink solitary waves can propagate stably under the additive white noise
perturbation. Next we study the evolution of the chirped periodic waves in
the negative index material in the presence of small initial perturbations.
Here we take the chirped periodic wave (\ref{33}) as an example to analyze
the stability of the solution numerically. We performed direct simulations
with amplitude perturbation and initial white noise to study the stability
of the solution (\ref{33}) compared to Fig. 2(b). The numerical results are
shown in Fig. 10(a) in which the initial amplitude is perturbed and Fig.
10(b) in which the solution is affected by the random noise. The results
reveal that the finite initial perturbations of amplitude and additive white
noise could not influence the main character of the periodic solutions.
Therefore, the chirped solitary and periodic waves show structural stability
with respect to the small input profile perturbations. Thus we conclude that
the nonlinearly chirped structures we presented are stable in an appropriate
range of parameters for soliton and periodic solutions of the generalized
NLSE with pseudo-quintic nonlinearity and self-steepening effect.

\section{Conclusion}

In conclusion, we have demonstrated that several new fascinating types of
periodic waves are formed in a NIM under the influence of pseudo-quintic
nonlinearity and self-steepening effect. Such structures exhibit an
interesting chirping property which shows a dependence on the light field
intensity. Remarkably, the nonlinearity in pulse chirp is found to be caused
by the presence of self-steepening process in the negative refractive index
material. This is contrary to the case of vanishing self-steepening effect
where only the unchirped envelope pulses are allowed to exist in the system
in the presence of pseudo-quintic nonlinearity. Chirped periodic wave
solutions expressed in terms of trigonometric functions have been also
obtained for the governing generalized NLSE. These results constitute the
first analytical demonstration of existence of periodic waves having a
nonlinear chirp in negative-index media. Considering the long-wave limit of
the derived periodic waves, we obtained a diversity of chirped localized
solutions including bright, dark, kink, anti-kink, and gray solitary pulses.
We further studied the stability of the nonlinearly chirped structures by
using the theory of nonlinear dispersive waves and direct numerical
simulations. Our results based on stability criterion and numerical
simulations showed the robustness of those wave forms with respect to finite
perturbations of the amplitude and additive white noise, thus motivating the
experimental observations of such structures in materials which have
negative index of refraction. Hence, we conclude that localized and periodic
waves exhibiting a nonlinear chirp can be formed in negative refractive
index media under the influence of pseudo-quintic nonlinearity and
self-steepening effect. It is predicted in this paper that such chirped
waves are experimentally realizable and hence these structures should be of
relevance for applications requiring the use of a negative index material.
One should note that the use of the above-mentioned generalized NLSE is not
only restricted to ultrashort electromagnetic pulse propagation in negative
refractive index media, but also to the description of light pulse dynamics
in nonlinear fibers. Therefore the results presented in this work are also
helpful for understanding the behavior of nonlinear waves in optical fiber
systems. In the latter setting, the chirping property is of great advantage
in fiber-optic applications such as the compression and amplification of
light pulses and thus chirped pulses are particularly useful in the design
of fiber-optic amplifiers, optical compressors, and solitary-wave-based
communications links.

\end{document}